\documentclass[
 reprint,superscriptaddress,
 amsmath,amssymb,
 aps]{revtex4-1}

\usepackage{graphicx}
\usepackage{dcolumn}
\usepackage{bm}
\usepackage[ruled]{algorithm2e}
\usepackage{multirow,booktabs}
\usepackage{svg}
\usepackage{paralist,siunitx,soul}
\usepackage[version=4]{mhchem}
\usepackage{hyperref}
\usepackage{url}
\hypersetup{colorlinks,urlcolor=blue,citecolor=blue,linkcolor=blue}
\DeclareUrlCommand\url{\color{blue}}

\graphicspath{{figs/}}

\begin{document}

\title{Graph deep learning accelerated efficient crystal structure search \\ and feature extraction}

\author{Chuannan Li}
\thanks{These authors contributed equally to this work.}
\affiliation{Hefei National Laboratory, Department of Physics, University of Science and Technology of China, Hefei, Anhui 230026, China}

\author{Hanpu Liang}
\thanks{These authors contributed equally to this work.}
\affiliation{Beijing Computational Science Research Center, Beijing, 100193, China}

\author{Xie Zhang}
\affiliation{Beijing Computational Science Research Center, Beijing, 100193, China}

\author{Zijing Lin}
\email{zjlin@ustc.edu.cn}
\affiliation{Hefei National Laboratory, Department of Physics, University of Science and Technology of China, Hefei, Anhui 230026, China}

\author{Su-Huai Wei}
\email{suhuaiwei@csrc.ac.cn}
\affiliation{Beijing Computational Science Research Center, Beijing, 100193, China}


\begin{abstract}
Structural search and feature extraction are a central subject in modern materials design, the efficiency of which is currently limited, but can be potentially boosted by machine learning (ML).
Here, we develop an ML-based prediction-analysis framework, which includes a symmetry-based combinatorial crystal optimization program (SCCOP) and a feature additive attribution model, to significantly reduce computational costs and to extract property-related structural features.
Our method is highly accurate and predictive, and extracts structural features from desired structures to guide materials design.
As a case study, we apply our new approach to a two-dimensional B-C-N system, which identifies 28 previously undiscovered stable structures out of 82 compositions; our analysis further establishes the structural features that contribute most to energy and bandgap.
Compared to conventional approaches, SCCOP is about 10 times faster while maintaining a comparable accuracy.
Our new framework is generally applicable to all types of systems for precise and efficient structural search, providing new insights into the relationship between ML-extracted structural features and physical properties.

\end{abstract}

\maketitle

\section{Introduction}

Predicting the crystal structure for a given composition prior to experimental syntheses is central to computation-guided materials discovery.
The state-of-the-art approaches for crystal structure prediction rely on efficient search algorithms such as simulated annealing (SA) \cite{SA_1,SA_2,SA_3}, genetic algorithm (GA) \cite{GA_1,GA_2,GA_3}, and particle-swarm optimization (PSO) \cite{PSO_1,PSO_2,CALYPSO_2D}.
These approaches require, however, extensive energy and force evaluation by density functional theory (DFT) \cite{DFT_1,DFT_2}  when exploring the configuration space.
As the numbers of atoms and species increase, the number of configurations grows exponentially, leading to an intolerable time and resources consumption.
In this context, machine learning (ML) is particularly powerful in reducing the computational consumption by adopting a surrogate model, e.g., crystal graph convolutional neural network (CGCNN) \cite{CGCNN}, and other graph-based prediction models \cite{ALIGNN,MEGNet,MPNN}. 
For instance, CGCNN considers the crystal topology to build undirected multigraphs, which can efficiently integrate structural features and be used to predict physical properties to replace DFT calculations.

After a large amount of structural searches, extracting the property-related structural features is essential for the exploration of new materials.
To deeply explore and visualize the underlying relationship between global and local atomic structures and physical properties such as stability and conductivity, numerous efforts have been made \cite{CGCNN_XAI,ALIGNN-d}.
For example, the transformation between fold and unfold states in protein-folding dynamics has been unveiled by encoding the entire mapping from biomolecular coordinates to a Markov state model \cite{VAMPnet};
similarly, the transition that contributes to Li-ion conduction can also be clearly verified by using graph dynamical network to learn local atomic environment instead of global dynamics \cite{GDNet}.
These studies imply that local atomic-scale structural motifs play a critical role in physical properties. 
However, this relationship still remains unclear in the structural generation field because of huge possible materials population and complex interatomic bonding, which are difficult to analyze by conventional methods.
An ML-based framework for structural search and data analysis is thus in critical demand in order to improve the efficiency of exploring new materials.

Two-dimensional (2D) materials are under extensive research, especially after the successful syntheses of novel 2D materials such as carbon biphenylene \cite{Biphenylene1} and T-carbon nanowires \cite{2011-PRL-Tcarbon,2017-NC-T-carbon-exp}, for fancinating physical phenomena induced by special structural features, e.g., nonhexagonal bonding and carbon tetrahedron. 
Since the differences in atomic mass and electronegativity are small enough, boron, carbon and nitrogen elements can be combined into abundant planar \ce{B_xC_yN_{1-x-y}} compounds \cite{BCN_1,Biphenylene2,Liang_1} and enable the flexibility to modulate stability and electronic structure by tuning the alloy composition \cite{CN_Study}. 
Nevertheless, systematic structural searches for the B-C-N alloy system are still rare \cite{BC_PSO_Search,BN_PSO_Search}.

\begin{figure*}[!htb]
	\includegraphics[width=16.5cm]{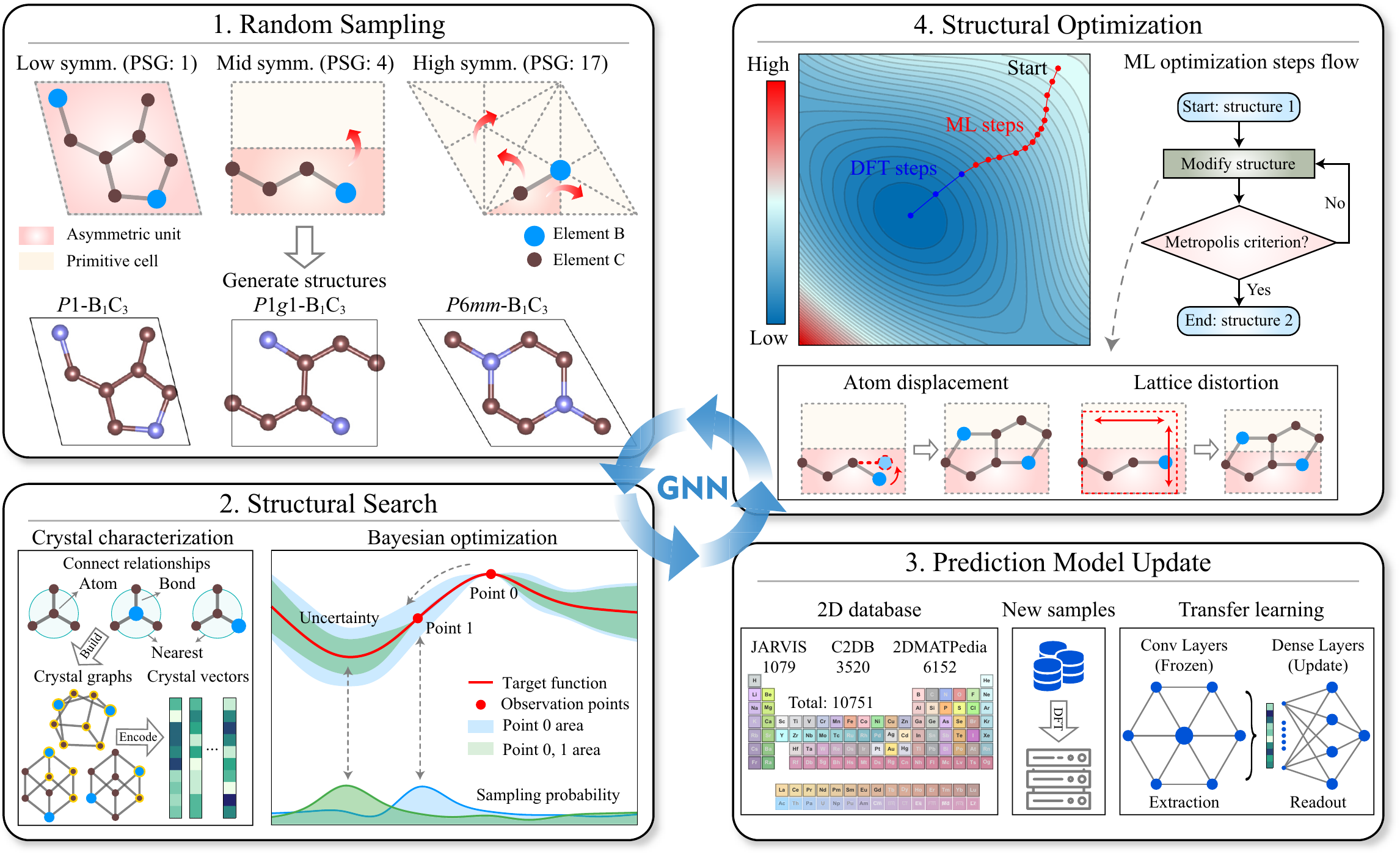}
	\caption{\label{fig:fig1} Workflow of SCCOP for the search of two-dimensional materials. 1) Generating structures by symmetry; 2) characterizing structures into crystal vectors and exploring the potential energy surface by a Bayesian optimization; 3) updating energy prediction model; 4) optimizing structures to obtain the lowest-energy configuration by ML and DFT. The whole program runs in a closed loop.}
\end{figure*}

In this work, we construct a prediction-analysis framework that combines a symmetry-based combinatorial crystal optimization program (SCCOP) for structural search of target compositions and a feature additive attribution model for data analyses.
A practical demonstration is performed for 2D B-C-N system to illustrate the high-throughput structural search and the ability on extracting structural features.
We first convert the structures generated from 17 plane space groups to crystal vectors by graph neural network (GNN) and predict their energies.
A Bayesian optimization is performed to explore the structure at the minimum of the potential energy surface.
For the desired structures, we optimize it with ML-accelerated SA, in conjunction with a limited number of DFT calculations to obtain the lowest energy configuration.
We further demonstrate that the additive feature attribution model can efficiently capture the structural features that dominate the energy and bandgap.
We identify five novel low-energy semiconductors among all the B-C-N compounds, which have bandgaps and mechanical performance comparable with 2D hexagonal BN.
Finally, we compare the performance of three methods: SCCOP, DFT-GA and DFT-PSO, which indicates that SCCOP is about 10 times faster while maintaining comparable accuracy.

\section{Methods}

The framework of prediction-analysis consists of five parts: i) random sampling, ii) structural search, iii) prediction model update, iv) structural optimization, and v) structural analysis.
The workflow of SCCOP is illustrated in Fig.~\ref{fig:fig1}, where GNN characterizes the crystal structures and connects each part to achieve iterations.

\textbf{Random Sampling.}
In the first step of SCCOP, to roughly measure the potential energy surface, unbiased initial structures are randomly generated from 17 plane space groups (PSGs), which cover all types of symmetry of 2D materials, as shown by step 1 in Fig.~\ref{fig:fig1}.
To determine the structure with a target composition, only the periodic lattice $ \bm{L}=(\bm{l}_{1},\bm{l}_{2},\bm{l}_{3})\in\mathbb{R}^{3\times 3}$, PSG, atom types $ \bm{A}=(a_{0},...,a_{N}) $, and atomic positions $ \bm{X}=(\bm{x}_{0},...,\bm{x}_{N})\in\mathbb{R}^{N\times 3} $ are necessary.
The $ n $ atoms of a structure are placed in an asymmetric unit (AU) \cite{Wyckoff}, which is the irreducible space and can fill the primitive cell by applying symmetry operations, enabling efficient configurational evolution.  
The space discretization and minimal interatomic distance techniques \cite{Gridding} are employed to reduce the search space.
A set of reasonable crystal structures $ \mathcal{C} $ can thus be generated efficiently.
All asymmetric units used in SCCOP are listed in Tables S1-S3.

\textbf{Structural Search.}
To further constrain the search space, a Bayesian optimization is applied to redistribute the sampling probability in order to find the energetically favorable structures, as illustrated in step 2 in Fig.~\ref{fig:fig1}.
In this step, crystal structures are first converted to crystal vector $ \bm{c} $ to achieve crystal characterization.
A crystal graph $ \mathcal{G} $ is built upon the atoms in AU to maximize the efficiency of GNN (Table S4), and the graph convolutional operator \cite{CGCNN} defined as $\bm{v}^{(t+1)}_{i}=\mathrm{Conv}(\bm{v}_{i}^{(t)},\bm{v}_{j}^{(t)},\bm{u}_{(i,j)_{k}}) $, where $ \bm{v}_{i}^{(t)},\bm{v}_{j}^{(t)} $ and $ \bm{u}_{(i,j)_{k}} $ are atom feature vectors and bond feature vectors at $ t $ convolution, respectively.
After $ K $ convolutions, the crystal vector $ \bm{c}=\bm{W}_{m}\bm{V} $ is the weighted sum of atom vectors $ \bm{V}=(\bm{v}^{(K)}_{1},...,\bm{v}^{(K)}_{n})\in\mathbb{R}^{n\times 64} $, where $ \bm{W}_{m}=(w_{1},...,w_{n}) $ denotes the multiplicity weight matrix that depends on the symmetry of atoms.
Lastly, two dense layers are added to map crystal vector $ \bm{c} $ to $ \hat{E} $; hence, a rough energy estimation of structures in $ \mathcal{C} $ can be realized by the GNN model.
A few low-$ \hat{E} $ structures are selected to obtain more precise energies by DFT calculations for the Bayesian optimization.

Approximating the function $ E=U(\bm{c}) $ between energy and structures is key for the Bayesian optimization.
Here we characterize the structures by the crystal vectors and use samples from precise DFT calculations to fit the function $ U $ by a Gaussian Process Model \cite{Gaussian_Model}.
The probability of improvement \cite{Bayesian} is adopted as the acquisition function $ PI(\bm{c})=1-\Phi((\mu(\bm{c})-U(\bm{c}^{*})-\xi)/\sigma(\bm{c})) $, where $ \bm{c}^{*}=\mathop{\arg\min}_{i} U(\bm{c}_{i}) $;
$ \mu(\bm{c}) $ and $ \sigma(\bm{c}) $ are the mean and standard deviations of the posterior distribution on $ \bm{c} $ from the Gaussian Process, respectively, and $ \Phi $ is the cumulative distribution function for a normal distribution.
The $ \xi $ parameter is used to balance the trade-off between exploitation and exploration.
We calculate $ PI $ among $ \mathcal{C} $, and choose high-acquisition-value structures for further structural optimization.

\textbf{Prediction Model Update.}
For target compositions, the pretrained GNN prediction model should be slightly updated to reach a better accuracy, as seen in step 3 in Fig.~\ref{fig:fig1}.
The pretrained model is trained by the 2D material databases JARVIS-DFT \cite{JARVIS}, C2DB \cite{C2DB}, and 2DMATPedia \cite{2DMatPedia}, which contain 10751 crystals covering 85 elements, 4 lattice systems and 17 PSGs.
The train:validation:test ratio is 60\%:20\%:20\%; a batch of 128 structures with the Adam optimizer \cite{Adam} is used, and the best-performing model in validation set is chosen as the pretrained model.
The lowest mean absolute error (MAE) in the validation set is 0.1468 eV/atom, with a smaller MAE of 0.1451 eV/atom in the test set, implying that the model has a strong generalization ability (shown in Fig. S1).
According to the transfer learning techniques \cite{Transfer_Learning}, when a small amount of DFT data is used in the search, the prediction model freezes the parameters of graph convolutional layers and only optimizes the full connected layers, which prevents overfitting of the DFT data and improves the capability of distinguishing the energy changes for different predicted structures.

\textbf{Structural Optimization.}
To obtain more accurate structural parameters and energies of target structures, SCCOP optimizes the structures by first ML and then DFT , as illustrated in step 4 in Fig.~\ref{fig:fig1}.
The structures occupy the relatively high-energy area on the potential energy surface. 
We first optimize the structural candidates with the ML-accelerated SA.
ML adjusts the structures by displacing the atomic positions and distorting lattice vectors with the Metropolis criterion \cite{SA_1}, i.e., using the probability $ \exp(-\Delta \hat{E}/kT) $ to decide if the changes are accepted according to the energy differences $ \Delta \hat{E} $ given by the GNN prediction model.
For the ML-optimized structures, $ t $-distributed stochastic neighbor embedding (TSNE) \cite{TSNE} is performed to reduce the dimension of crystal vectors and the Kmeans method \cite{Kmeans} is used to group the vectors into different clusters.
Then DFT optimization is performed to more rigorously relax the structure (that has the lowest energy in each cluster) to find the local minimum on the potential energy surface.
The optimized lattice in this step will be employed as the initial lattice in the next search iteration to sample new crystal structures.

\textbf{Structural Analysis.}
An additive feature attribution model \cite{CGCNN_XAI,XAI_Add_Model_Review} is applied to extract property-related features from massive amounts of data. 
Thus, the averaged total energy per atom is predicted by the sum over different local chemical environments, i.e., $ \hat{E}=\sum_{i}^{N} \hat{E}_{i}/N $, where $ \hat{E}_{i}=\bm{W}_{l}\bm{v}^{T}_{i}+b_{l} $ is built by the atom feature vector $\bm{v}^{T}_{i}$, the weight $\bm{W}_{l}$, and the bias $ b_{l} $.
To focus on the environment consisting of center and neighbor atoms, we calculate its contribution to energy $ \bar{E}_{i} $ by the average of $ \hat{E}_{i} $ on the data that are clustered by coordination atoms, bond lengths, and bond angles. 
In this way, the energy contribution from each structural motif can be accessed independently, and lower $ \bar{E} $ means higher local structural stability.
Meanwhile, for solid-solution systems, the bandgap $ \hat{G}=\sum_{i}^{N} \hat{G}_{i}/N $ is analyzed in the same way.
$ \hat{G}_{i} $ is also calculated by a linear transformation acting on $ \bm{v}^{T}_{i} $, with a specifically designed loss function $ \mathcal{L}=\hat{\mathbb{E}}_{G>0}[(G-\hat{G})^{2}]+\hat{\mathbb{E}}_{G=0}[(G-\max(\hat{G}, 0))^{2}] $;
the expectation $ \hat{\mathbb{E}}[...] $ indicates an average over a finite batch of samples, and $ G $ is the bandgap computed from DFT.
Therefore, structures with zero or negative $ \hat{G} $ are classified as metal, which makes $ \hat{G}_{i} $ a physically meaningful term; a positive $ \hat{G}_{i} $ means opening the bandgap, otherwise closing the bandgap.
Both of the two analysis models are trained with 80\% of the data and then validated with the remaining 20\% of the data; the best-performing model in the validation set is selected.

\textbf{DFT Calculations.}
The DFT relaxations, energy and bandgap calculations for the searched structures are carried out using the Vienna Ab-initio Simulation Package (VASP) \cite{VASP_1,VASP_2,VASP_3}.
For structural relaxations and energy evaluations, the generalized gradient approximation (GGA) within the Perdew-Burke-Ernzerhof (PBE) form for the exchange-correlation functional \cite{PBE} is used.
The ion-electron interactions are treated by projector-augmented-wave (PAW) \cite{PAW_1,PAW_2} technique.
The plane-wave energy cutoff is set to 520 eV.
The Brillouin zone associated with the primitive cell is sampled using a Monkhorst-Pack $ k $-point mesh of $4\times4\times1$.
A vacuum space of 15 $ \mathrm{\AA} $ is applied to avoid artificial interactions between the periodic images.
All structures are relaxed with energies and forces converged to $ 10^{-5} $ eV and $ 0.01 $ eV/$ \mathrm{\AA} $, respectively. 
The electronic band structures are calculated with the HSE06 hybrid functional \cite{HSE06}.
The phonon thermal conductivity is predicted by the ShengBTE code \cite{ShengBTE}.

\begin{figure*}[!htb]
	\includegraphics[width=16cm,trim=0 0 0 55,clip]{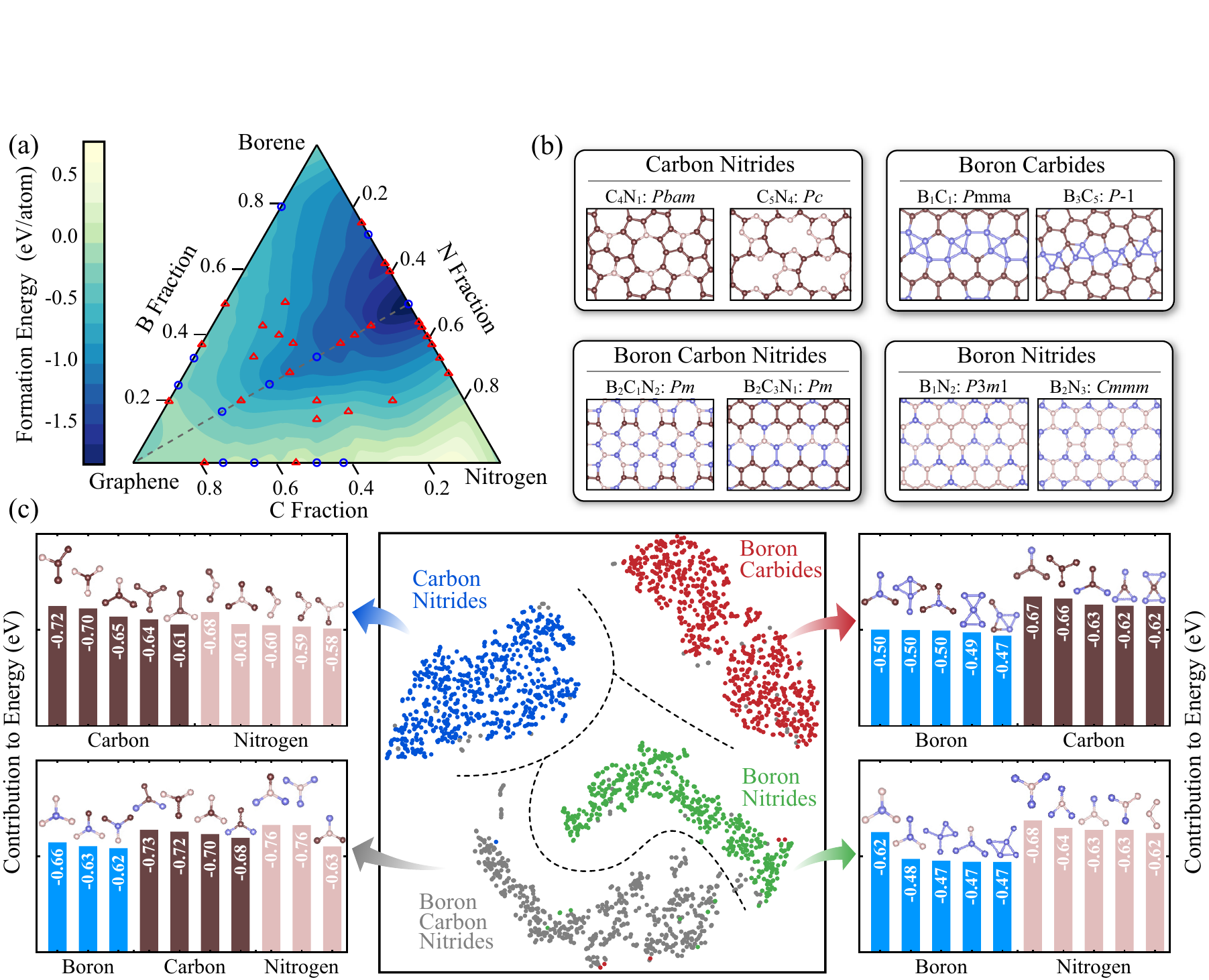}
	\caption{\label{fig:fig2} (a) Ternary phase diagram of the B-C-N system. All calculations are carried out at 0 K. The borene, graphene, and nitrogen are chosen as the corners of the Gibbs triangle. Blue circles and red hollow triangles represent stable compounds and newly found stable structures, respectively; the gray dashed line indicates the compositions with a B:N ratio of 1. (b) Illustration of typical stable structures of four compounds searched by SCCOP. (c) Distribution of two-dimensional crystal vectors on a 2D plane using the TSNE dimensionality reduction. Energy contribution of the structural motifs in four compounds are listed on the sides; each motif contains center and neighbor atoms.}
\end{figure*}

\section{\label{sec:level3}Results and Discussion}

We employ SCCOP to explore 82 different compositions of the B-C-N system (see Figs. S2-S5); for each composition, we select the structures up to 0.5 eV/atom above the convex hull, and a total of 2623 structures are identified.
Further, we analyze the average energy and bandgaps with structural features extracted by the additive feature attribution model.
By these approaches, five N-rich wide-bandgap insulators are newly discovered. Lastly, we compare SCCOP with other DFT-based methods, such as DFT-GA and DFT-PSO that have been employed in the mainstream USPEX \cite{GA_3} and CALYPSO \cite{PSO_2} structural search codes, respectively.

\begin{figure*}[!htb]
	\includegraphics[width=16cm]{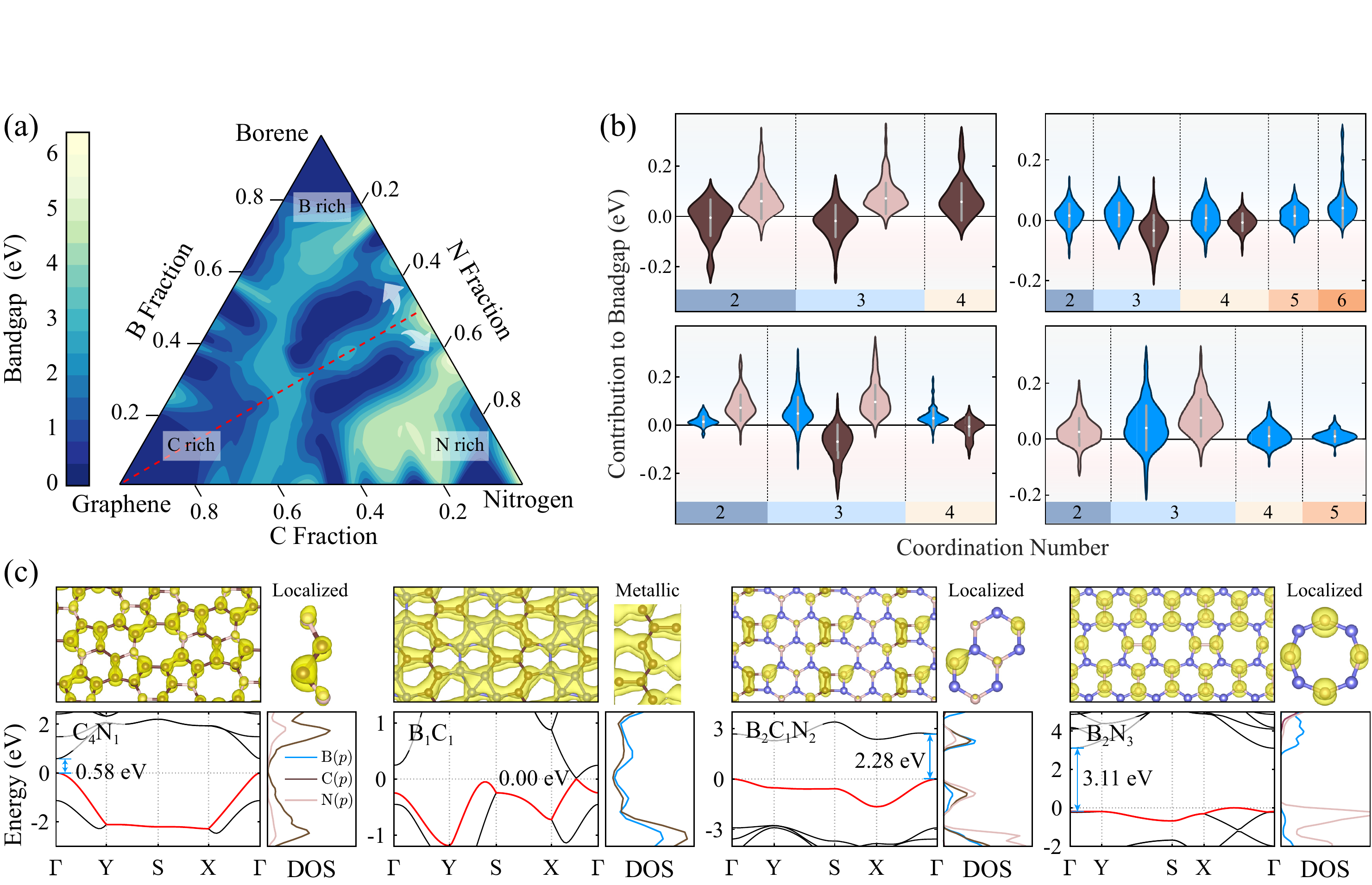}
	\caption{\label{fig:fig3} (a) Bandgap distribution of the B-C-N system. For each composition, the bandgap of the lowest-energy structure is calculated. The red dashed line indicates the compositions with a B:N ratio of 1. (b) Contributions to the bandgap from different coordination numbers. Brown, pink and blue colors denote carbon, nitrogen and boron, respectively. (c) Structural features for opening or closing the bandgap of 4 typical structures; the spatial valence band edge state distribution, band structure near the Fermi level as well as the density of state (DOS) are also depicted. The bandgap contributions and structural features are obtained from the additive feature attribution model.}
\end{figure*}

\subsection{Energy-Related Feature Extraction}
For a thorough understanding of the connection between stability and structural feature, we first plot the ternary phase diagram of the B-C-N system in Fig.~\ref{fig:fig2}(a).
In addition to 11 previously reported structures (blue circle) \cite{CN_Review,BCN_Review_1,BCN_Review_2}, 28 dynamically stable low-energy structures are newly discovered (red hollow triangle), e.g., \ce{B1C1}, \ce{B1N2}, \ce{C4N1}, and \ce{B2C1N2} [listed in Fig.~\ref{fig:fig2}(b)].
The stable phases of the B-C-N system have thus been greatly extended by the systematic search via SCCOP.
We note that the low-energy structures are located on a line, where the stoichiometric ratio of B:N is 1:1, e.g., \ce{BN}, \ce{BCN}, \ce{BC2N} and \ce{BC4N}, since the valence electrons of boron and nitrogen can be fully paired to reduce the energy of structure.
Similarly, the average valence electrons of boron carbides and carbon nitrides are either less or greater than four; they both hinder electrons pairing.
Thus, their formation energies are relatively high. 
The phonon spectra of all newly found stable structures are shown in Figs. S6-S8.

Next, we cluster structures by the crystal vectors and extract stable structural features in Fig.~\ref{fig:fig2}(c).
The crystal vectors strongly relate to the atomic species of the compounds and can be clearly grouped into four clusters: carbon nitrides (\ce{C_xN_{1-x}}), boron carbides (\ce{B_xC_{1-x}}), boron nitrides (\ce{B_xN_{1-x}}) and boron-carbon nitrides (\ce{B_xC_yN_{1-x-y}}).
This indicates that the compounds in the same cluster have similar electronic structures to form structural features with similar energies, making it possible for GNN to predict energy from these features.
For all four compounds, ML finds that $ sp^{2} $ hybridization with bond angles of $ 120^{\circ} $ is a universal structural feature, as the number of their valence electrons is close to four per atom.
The honeycomb structure might thus be energetically favorable.
In addition, the B-centered structural features contribute less to energy than those of carbon and nitrogen.
This is primarily due to its electron-deficient bonding nature \cite{Boron_Deficient}.
Carbon and nitrogen atoms can, however, form conjugated $ \pi $ bonds or fill empty $ p $ orbitals with lone pairs of electrons to enhance the stability.
In the carbon nitrides, two most common types of nitrogen atoms are found, i.e., pyridinic-N (-0.68 eV) and graphitic-N (-0.61 eV) \cite{CN_Review}.
For pyridinic-N, the nitrogen atom is coordinated to two carbons and one orbital is occupied by a lone-pair of electrons, while graphitic-N is characterized by nitrogen $ sp^2 $ hybridization with three carbon atoms.
In the boron carbides and boron nitrides, the boron atoms tend to bond with more than three atoms, implying that boron can stabilize the structure by forming coordination bonds or multi-centered bonds \cite{BC_PSO_Search}.
Moreover, because of the good match on the chemical valence, three-fold coordination dominates the structural features of boron carbon nitrides.
These extracted structural features deepen the understanding of structural stability and may guide future searches of low-energy B-C-N materials.

\begin{figure*}[!htb]
	\includegraphics[width=16cm]{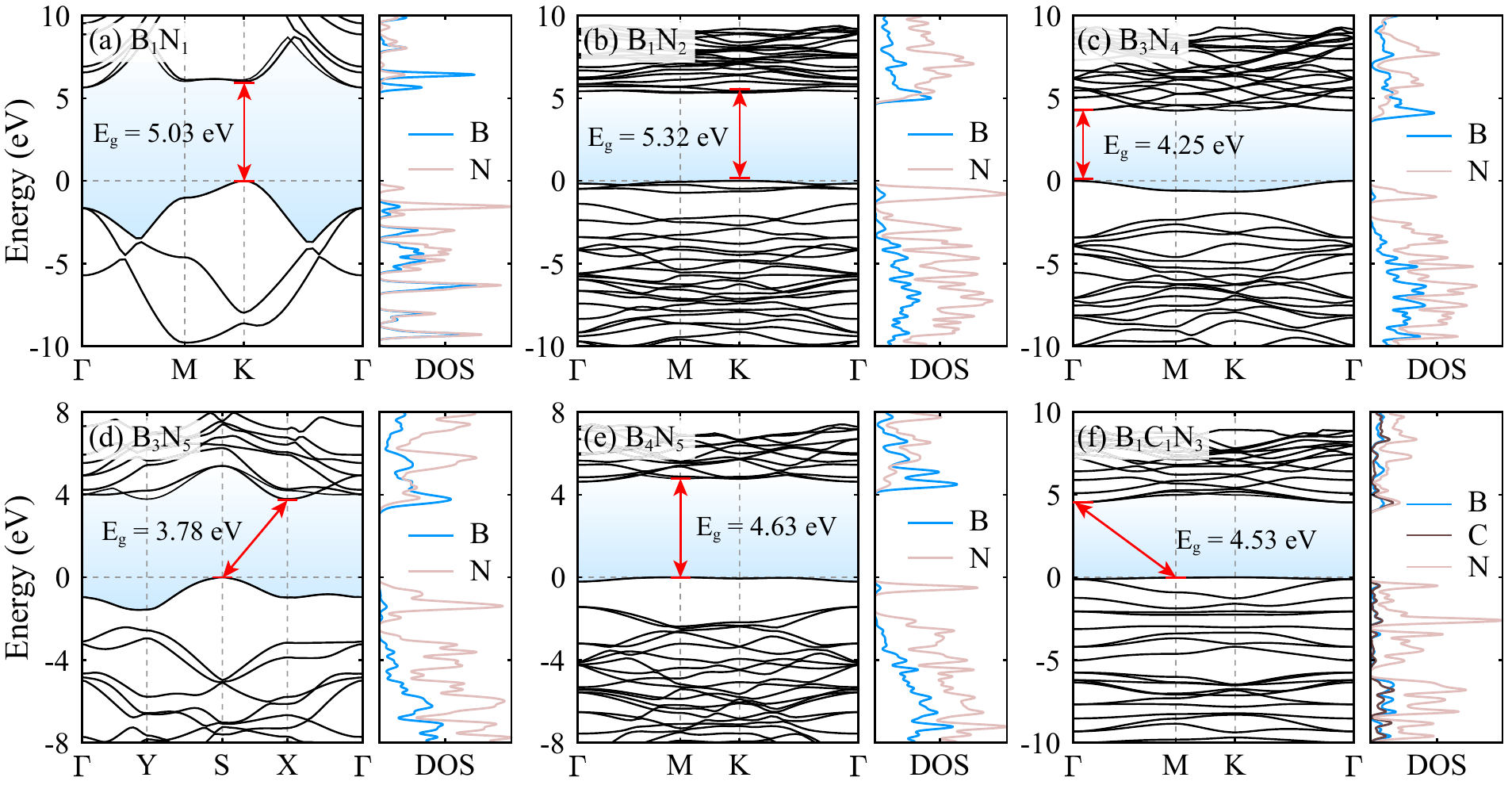}
	\caption{\label{fig:fig4} Electronic band structures and density of states (DOS) for (a) $h$-BN and (b)-(f) the newly discovered wide-bandgap materials.}
\end{figure*}

\subsection{Bandgap-Related Feature Extraction}
To find out how element composition and bandgap are related, the bandgap distribution of the B-C-N system is plotted in Fig.~\ref{fig:fig3}(a), which shows narrower bandgaps for the B-rich and C-rich compositions and wider bandgaps for the N-rich compositions.
Interestingly, two metallic phase regimes are located on two sides of a line with a B:N ratio of 1 (see the red dashed line in Fig.~\ref{fig:fig3}); this is because the mismatch of valence electrons, which form a band crossing the Fermi level. 
Suitable compositions (e.g., B:N=3:1 and 1:3) help to open the band gap, while the N-rich compounds are more likely to have larger bandgaps.
We cluster the structural features by the coordination number in Fig.~\ref{fig:fig3}(b).
2-fold and 3-fold coordination carbon atoms play a key role in closing the bandgap due to the free $ p $ electrons.
However, 4-fold coordination carbon, strong electronegative nitrogen, 6-fold coordination boron have little contributions to the electrical conductivity due to either fully paired of electrons or absence of free electrons.
Overall, ML enables the bandgap analysis from the perspective of coordination number, allowing to draw conclusions that are consistent with our physical intuition.

Furthermore, we consider the contribution of larger structural features comprising several atoms to the bandgap.
The percentage of contribution is defined by $ F=\sum_{i}G_{i}/G_{\mathrm{tot}}\times 100\% $, where the summation is over the atoms in the selected structural feature and $ G_{\mathrm{tot}} $ is the total contribution to open or close the bandgap.
Therefore, greater $ F $ implies that this structural feature is more important to the bandgap.
Here, four structures are given as examples to show the main factor identified by ML that relates to the formation of bandgap in Fig.~\ref{fig:fig3}(c).
In \ce{C4N1}, the band-edge states are mainly projected on the N-C-C-N chain, and ML identifies that the chain provides 86\% contribution to the band-edge states.
The N-C-C-N chain introduces a localized low-energy impurity energy level near the Fermi level, thus leading to the split of electron cloud in 5-, 6-, and 8-membered rings.
In \ce{B1C1}, C chains are identified to be the central factor in closing the bandgap (100\% contribution), which enable the formation of continuous electron clouds and spread to the empty orbitals of adjacent boron atoms.

\begin{table}[tp]
	\caption{\label{tab:table1} Calculated Young's modulus ($ E $), Poisson's ratio ($ \nu $), shear moduli ($ G $), and lattice thermal conductivity ($ \kappa $) at 300K for $h$-BN (\ce{B1N1}) and the discovered wide-bandgap materials.}
	\begin{ruledtabular}
		\renewcommand{\arraystretch}{1.1}
		\begin{tabular}{lcccc}
			Structures & $ E $ (N/m) & $ \nu $ (N/m) & $ G $ (N/m) & $ \kappa $ (W/mK)\\ \hline\rule{-3pt}{10pt}
			\ce{B1N1} & 185.92 & 0.22 & 76.50 & 708.07\\
			\ce{B1N2} & 132.83 & 0.08 & 71.70 & 10.13\\
			\ce{B3N4} & 180.24 & 0.19 & 75.90 & 65.21\\
			\ce{B3N5} & 179.50 & 0.17 & 76.05 & 60.00\\
			\ce{B4N5} & 172.41 & 0.16 & 74.35 & 41.75\\
			\ce{B1C1N3} & 113.60 & 0.24 & 45.84 & 31.62
		\end{tabular}
	\end{ruledtabular}
\end{table}

In \ce{B2C1N2} and \ce{B2N3}, 6- and 8-membered rings of alternating B-N bonds contribute 100\% and 75\% to the band-edge states to enlarge the bandgap, respectively.
Both of them are formed by the same structural motif that is characterized by nitrogen coordination with boron atoms with electrons localized on nitrogen,
The direct wide-bandgap insulator hexagonal BN ($h$-\ce{BN}) is composed entirely of this feature.
In general, ML can quantify the contribution percentage for a given structural feature to rationalize the formation of bandgap. 
However, the selection of multi-atom structural features still requires human assistance to verify the rationality; a general method for the selection of features is still in demand.

\subsection{Wide-bandgap Insulators}

\begin{figure*}[!htb]
	\includegraphics[width=17cm]{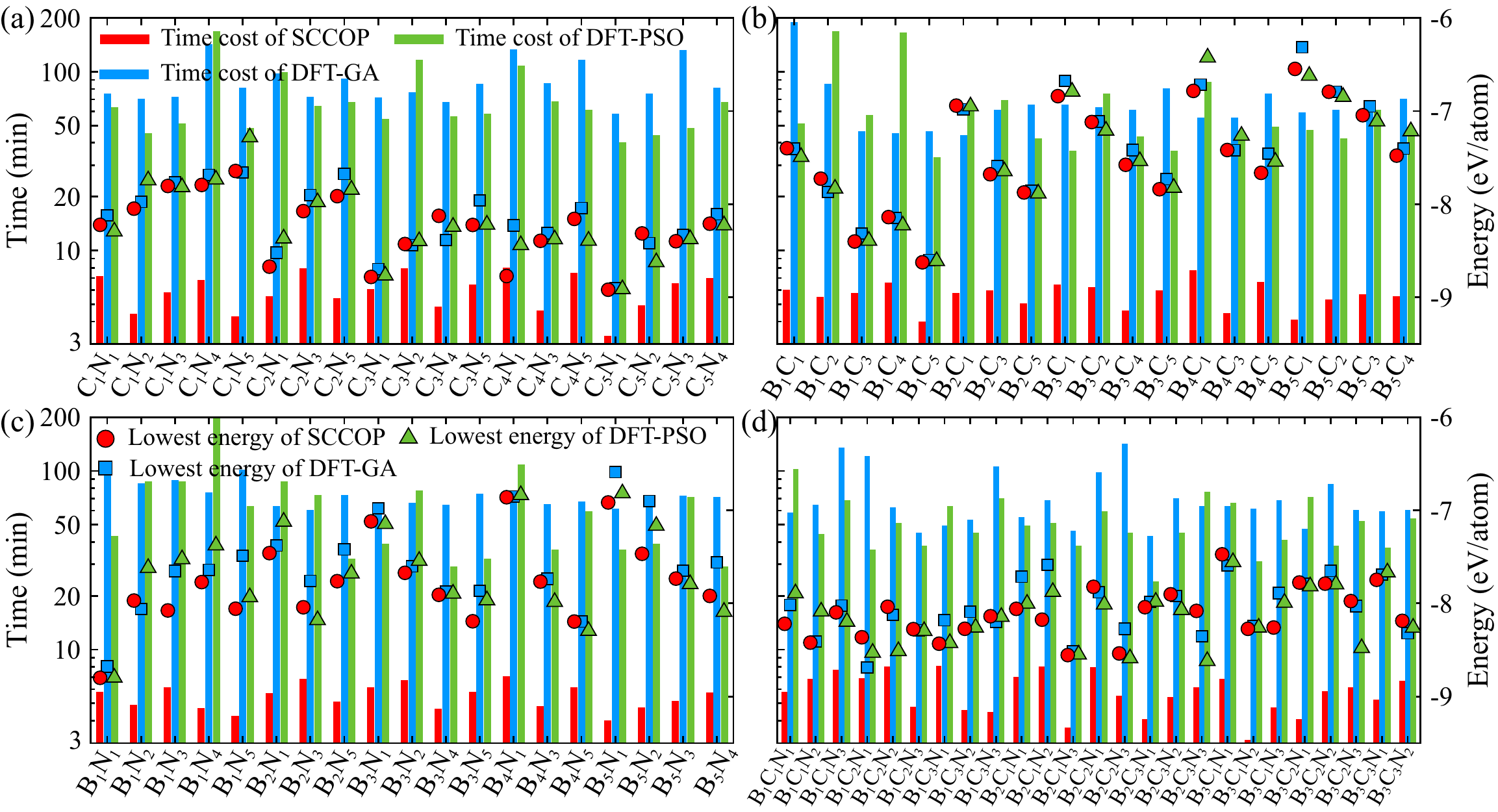}
	\caption{\label{fig:fig5}  Comparison of computational time cost and the lowest energy found after 1 iteration in (a) carbon nitrides, (b) boron carbides, (c) boron nitrides, and (d) boron carbon nitrides by SCCOP, DFT-GA, and DFT-PSO approaches. The left $y$-axis is the time cost in log scale and the right is the energy of the searched structures. The computational time is accounted for running on 2*GTX 1080 GPUs and 12*Xeon Gold 6248 CPUs.}
\end{figure*}

It is known to be challenging to predict N-rich materials, since two nitrogen atoms can easily combine into nitrogen molecule, resulting in ill structures during structural searches.
SCCOP solves the dilemma by quickly screening a large number of structures, with which we identify five stable wide-bandgap materials with bandgaps, mechanical performance, and structural motifs similar to $h$-BN in the N-rich area (see Figs.~\ref{fig:fig4}, S6, and S7, and Table~\ref{tab:table1}).
\ce{B1N2}, \ce{B3N4}, and \ce{B4N5} are direct-gap while \ce{B3N5} and \ce{B1C1N3} are indirect.
Especially, \ce{B1N2} has a bandgap (5.32 eV) that is even greater than that of $h$-\ce{BN}.
This is because the formation of the fully occupied N-$ p $ dangling-bond states reduces the hybridization and band width of the band-edge states, and thus enlarges the bandgap.
The Young's modulus, Poisson's ratio, and shear modulus of \ce{B3N4} are 180.24 N/m, 0.19 N/m, and 75.90 N/m, respectively.
The abundant strong bonding between boron and nitrogen in plane leads to the fact that \ce{B3N4} has comparable mechanical properties with $h$-\ce{BN}, and it is essential for the reliability in practical applications.
Moreover, the thermal conductivity of \ce{B1N2} is 10.13 W/mK, which is 70 times smaller than \ce{B1N1} (708.07 W/mK). 
The dramatic drop in the thermal conductivity is mainly caused by the asymmetric distribution of boron, carbon, and nitrogen atoms, which activates a phonon anharmonic effect, and hence results in the enhancement of phonon-phonon scattering to hinder thermal transport.
Overall, owing to the exotic optoelectronic properties, excellent mechanical robustness, and low thermal conductivity, the newly discovered materials thus have fruitful potential applications, e.g., in ultraviolet photodetectors \cite{Photodetectors1, Photodetectors2}, thermal insulation materials \cite{ThermalInsulator1, ThermalInsulator2}, and energy storage devices \cite{Storage1, Storage2}.

\subsection{Method Comparison}

\begin{table}[tp]
	\caption{\label{tab:table2} Comparison of the time cost and successful rate of three structural search methods for class of four compounds, where the successful rate is the ratio that the method finds the lowest-energy structures in three methods. All methods run 1 iteration for comparison.}
	\begin{ruledtabular}
		\renewcommand{\arraystretch}{1.1}
		\begin{tabular}{lccc}
			Compounds & Methods & Successful Rate & Time (min) \\  \hline \rule{-3pt}{10pt}
			\ce{B_xC_{1-x}} & DFT-GA & 3/19 & 67.8 \\
			& DFT-PSO & 9/19 & 63.7 \\
			& SCCOP & \textbf{10/19} & \textbf{5.6} \\ \hline \rule{-3pt}{10pt}
			\ce{B_xN_{1-x}} & DFT-GA & 1/19 & 73.2 \\
			& DFT-PSO & 7/19 & 64.5 \\ 
			& SCCOP & \textbf{12/19} & \textbf{5.5} \\ \hline \rule{-3pt}{10pt}
			\ce{C_xN_{1-x}} & DFT-GA & 3/19 & 88.3 \\
			& DFT-PSO & 8/19 & 69.7 \\
			& SCCOP & \textbf{11/19} & \textbf{6.0} \\ \hline \rule{-3pt}{10pt}
			\ce{B_xC_yN_{1-x-y}} & DFT-GA & 4/25 & 71.0 \\
			& DFT-PSO & 13/25 & 51.8 \\
			& SCCOP & \textbf{13/25} & \textbf{5.9} 
		\end{tabular}
	\end{ruledtabular}
\end{table}

Finally, we compare the computational performance of SCCOP with other commonly used DFT-based search approaches such as DFT-GA in USPEX and DFT-PSO in CALYPSO in Fig.~\ref{fig:fig4}.
All of them are tested on 82 compositions while ensuring that the parameter setup and computational resources were as consistent as possible.
Notably, SCCOP is the most time-saving among the three methods and performs well in most cases.
For a more concise understanding of the performance of the three methods, we summarize the key results of comparison in Table~\ref{tab:table2}.
We find that SCCOP identifies the lowest-energy structures among 45 compositions with an average time of 5.7 minutes, which is about 10 times faster than DFT-GA and DFT-PSO; the successful rate of SCCOP is comparable or even greater than that of the other two.
Therefore, we are confident that SCCOP can greatly reduce the search time while maintaining a comparable accuracy to the state-of-the-art DFT-based search approaches.
As a matter of fact, the GNN model is trained based on the DFT-calculated data; it thus cannot surpass the accuracy of DFT results.
However, due to the effective feature extraction and relative simple calculation style, GNN can predict energies faster than DFT by $ 3$--$5 $ orders of magnitude \cite{MPNN, GN_OA, BOWSR} while keeping a comparable accuracy.
Hence, the GNN-enhanced efficiency of SCCOP significantly reduces the time spent on initial structure screening and structural optimization, and this is the main reason why SCCOP can outperform DFT-based prediction methods.

\section{\label{sec:level4}Conclusions}

In summary, we have developed an ML-based framework for crystal structure prediction and analysis, which consists of five parts: i) generating abundant random structures in AU with symmetry and distance constraints, ii) Bayesian optimization with crystal graph representation for structures to search, iii) modifying the energy prediction model to adapt to target composition by transfer learning techniques, iv) carrying out GNN-accelerated SA for structural optimization, and v) constructing an additive feature attribution model for feature extraction of the search results.
We demonstrated this framework by applying it to predict the crystal structures of 82 compositions in the B-C-N system.
In addition to successful identification of previously unknown crystal structures, we were also able to extract the key features for structural stabilization, to establish the relationship between bandgap and coordination number, and to discover the critical factors for bandgap formation for specific structures.
Five stable wide-bandgap materials with excellent mechanical properties and low thermal conductivities have been successfully discovered.
Compared to conventional DFT-based prediction approaches and domain knowledge analysis methods, this integrated prediction-analysis framework, which takes full advantage of ML, can greatly shorten the discovery and design cycle of novel functional materials.

\hfill
\section{\label{sec:level6}Acknowledgments}

The work is sponsored by the National Natural Science Foundation of China (Nos. 12074362, 11774416, 52172136, 11991060, 12088101, and U1930402).
We also acknowledge computational resources from the Supercomputing Center of University of Science and Technology of China.

\section{\label{sec:level7}Author Contributions}

C.L. and H.L. equally contributed to developing the framework, preparing the figures and writing the manuscript. X.Z. contributed to the discussion of the results. Z.L. and S.-H.W. supervised and guided the project. All authors reviewed and edited the manuscript.

\section{\label{sec:level8}Competing Interests}

The authors declare no competing interests.

\nocite{*}

\bibliographystyle{apsrev4-2}

\begin{thebibliography}{60}%
	\makeatletter
	\providecommand \@ifxundefined [1]{%
		\@ifx{#1\undefined}
	}%
	\providecommand \@ifnum [1]{%
		\ifnum #1\expandafter \@firstoftwo
		\else \expandafter \@secondoftwo
		\fi
	}%
	\providecommand \@ifx [1]{%
		\ifx #1\expandafter \@firstoftwo
		\else \expandafter \@secondoftwo
		\fi
	}%
	\providecommand \natexlab [1]{#1}%
	\providecommand \enquote  [1]{``#1''}%
	\providecommand \bibnamefont  [1]{#1}%
	\providecommand \bibfnamefont [1]{#1}%
	\providecommand \citenamefont [1]{#1}%
	\providecommand \href@noop [0]{\@secondoftwo}%
	\providecommand \href [0]{\begingroup \@sanitize@url \@href}%
	\providecommand \@href[1]{\@@startlink{#1}\@@href}%
	\providecommand \@@href[1]{\endgroup#1\@@endlink}%
	\providecommand \@sanitize@url [0]{\catcode `\\12\catcode `\$12\catcode
		`\&12\catcode `\#12\catcode `\^12\catcode `\_12\catcode `\%12\relax}%
	\providecommand \@@startlink[1]{}%
	\providecommand \@@endlink[0]{}%
	\providecommand \url  [0]{\begingroup\@sanitize@url \@url }%
	\providecommand \@url [1]{\endgroup\@href {#1}{\urlprefix }}%
	\providecommand \urlprefix  [0]{URL }%
	\providecommand \Eprint [0]{\href }%
	\providecommand \doibase [0]{https://doi.org/}%
	\providecommand \selectlanguage [0]{\@gobble}%
	\providecommand \bibinfo  [0]{\@secondoftwo}%
	\providecommand \bibfield  [0]{\@secondoftwo}%
	\providecommand \translation [1]{[#1]}%
	\providecommand \BibitemOpen [0]{}%
	\providecommand \bibitemStop [0]{}%
	\providecommand \bibitemNoStop [0]{.\EOS\space}%
	\providecommand \EOS [0]{\spacefactor3000\relax}%
	\providecommand \BibitemShut  [1]{\csname bibitem#1\endcsname}%
	\let\auto@bib@innerbib\@empty
	\bibitem [{\citenamefont {Kirkpatrick}\ \emph {et~al.}(1983)\citenamefont
		{Kirkpatrick}, \citenamefont {Gelatt},\ and\ \citenamefont {Vecchi}}]{SA_1}%
	\BibitemOpen
	\bibfield  {author} {\bibinfo {author} {\bibfnamefont {S.}~\bibnamefont
			{Kirkpatrick}}, \bibinfo {author} {\bibfnamefont {C.~D.}\ \bibnamefont
			{Gelatt}},\ and\ \bibinfo {author} {\bibfnamefont {M.~P.}\ \bibnamefont
			{Vecchi}},\ }\href {https://doi.org/10.1126/science.220.4598.671} {\bibfield
		{journal} {\bibinfo  {journal} {Science}\ }\textbf {\bibinfo {volume}
			{220}},\ \bibinfo {pages} {671} (\bibinfo {year} {1983})}\BibitemShut
	{NoStop}%
	\bibitem [{\citenamefont {Wille}(1987)}]{SA_2}%
	\BibitemOpen
	\bibfield  {author} {\bibinfo {author} {\bibfnamefont {L.~T.}\ \bibnamefont
			{Wille}},\ }\href {https://doi.org/10.1038/325374c0} {\bibfield  {journal}
		{\bibinfo  {journal} {Nature}\ }\textbf {\bibinfo {volume} {325}},\ \bibinfo
		{pages} {374} (\bibinfo {year} {1987})}\BibitemShut {NoStop}%
	\bibitem [{\citenamefont {Doll}\ \emph {et~al.}(2007)\citenamefont {Doll},
		\citenamefont {Schön},\ and\ \citenamefont {Jansen}}]{SA_3}%
	\BibitemOpen
	\bibfield  {author} {\bibinfo {author} {\bibfnamefont {K.}~\bibnamefont
			{Doll}}, \bibinfo {author} {\bibfnamefont {J.~C.}\ \bibnamefont {Schön}},\
		and\ \bibinfo {author} {\bibfnamefont {M.}~\bibnamefont {Jansen}},\ }\href
	{https://doi.org/10.1039/B709943F} {\bibfield  {journal} {\bibinfo  {journal}
			{Phys. Chem. Chem. Phys.}\ }\textbf {\bibinfo {volume} {9}},\ \bibinfo
		{pages} {6128} (\bibinfo {year} {2007})}\BibitemShut {NoStop}%
	\bibitem [{\citenamefont {Deaven}\ and\ \citenamefont {Ho}(1995)}]{GA_1}%
	\BibitemOpen
	\bibfield  {author} {\bibinfo {author} {\bibfnamefont {D.~M.}\ \bibnamefont
			{Deaven}}\ and\ \bibinfo {author} {\bibfnamefont {K.~M.}\ \bibnamefont
			{Ho}},\ }\href {https://doi.org/10.1103/PhysRevLett.75.288} {\bibfield
		{journal} {\bibinfo  {journal} {Phys. Rev. Lett.}\ }\textbf {\bibinfo
			{volume} {75}},\ \bibinfo {pages} {288} (\bibinfo {year} {1995})}\BibitemShut
	{NoStop}%
	\bibitem [{\citenamefont {M.~Woodley}\ \emph {et~al.}(1999)\citenamefont
		{M.~Woodley}, \citenamefont {D.~Battle}, \citenamefont {D.~Gale},\ and\
		\citenamefont {Richard A.~Catlow}}]{GA_2}%
	\BibitemOpen
	\bibfield  {author} {\bibinfo {author} {\bibfnamefont {S.}~\bibnamefont
			{M.~Woodley}}, \bibinfo {author} {\bibfnamefont {P.}~\bibnamefont
			{D.~Battle}}, \bibinfo {author} {\bibfnamefont {J.}~\bibnamefont {D.~Gale}},\
		and\ \bibinfo {author} {\bibfnamefont {C.}~\bibnamefont {Richard
				A.~Catlow}},\ }\href {https://doi.org/10.1039/A901227C} {\bibfield  {journal}
		{\bibinfo  {journal} {Phys. Chem. Chem. Phys.}\ }\textbf {\bibinfo {volume}
			{1}},\ \bibinfo {pages} {2535} (\bibinfo {year} {1999})}\BibitemShut
	{NoStop}%
	\bibitem [{\citenamefont {Lyakhov}\ \emph {et~al.}(2013)\citenamefont
		{Lyakhov}, \citenamefont {Oganov}, \citenamefont {Stokes},\ and\
		\citenamefont {Zhu}}]{GA_3}%
	\BibitemOpen
	\bibfield  {author} {\bibinfo {author} {\bibfnamefont {A.~O.}\ \bibnamefont
			{Lyakhov}}, \bibinfo {author} {\bibfnamefont {A.~R.}\ \bibnamefont {Oganov}},
		\bibinfo {author} {\bibfnamefont {H.~T.}\ \bibnamefont {Stokes}},\ and\
		\bibinfo {author} {\bibfnamefont {Q.}~\bibnamefont {Zhu}},\ }\href
	{https://doi.org/https://doi.org/10.1016/j.cpc.2012.12.009} {\bibfield
		{journal} {\bibinfo  {journal} {Comput. Phys. Commun.}\ }\textbf {\bibinfo
			{volume} {184}},\ \bibinfo {pages} {1172} (\bibinfo {year}
		{2013})}\BibitemShut {NoStop}%
	\bibitem [{\citenamefont {Wang}\ \emph {et~al.}(2010)\citenamefont {Wang},
		\citenamefont {Lv}, \citenamefont {Zhu},\ and\ \citenamefont {Ma}}]{PSO_1}%
	\BibitemOpen
	\bibfield  {author} {\bibinfo {author} {\bibfnamefont {Y.}~\bibnamefont
			{Wang}}, \bibinfo {author} {\bibfnamefont {J.}~\bibnamefont {Lv}}, \bibinfo
		{author} {\bibfnamefont {L.}~\bibnamefont {Zhu}},\ and\ \bibinfo {author}
		{\bibfnamefont {Y.}~\bibnamefont {Ma}},\ }\href
	{https://doi.org/10.1103/PhysRevB.82.094116} {\bibfield  {journal} {\bibinfo
			{journal} {Phys. Rev. B}\ }\textbf {\bibinfo {volume} {82}},\ \bibinfo
		{pages} {094116} (\bibinfo {year} {2010})}\BibitemShut {NoStop}%
	\bibitem [{\citenamefont {Wang}\ \emph
		{et~al.}(2012{\natexlab{a}})\citenamefont {Wang}, \citenamefont {Lv},
		\citenamefont {Zhu},\ and\ \citenamefont {Ma}}]{PSO_2}%
	\BibitemOpen
	\bibfield  {author} {\bibinfo {author} {\bibfnamefont {Y.}~\bibnamefont
			{Wang}}, \bibinfo {author} {\bibfnamefont {J.}~\bibnamefont {Lv}}, \bibinfo
		{author} {\bibfnamefont {L.}~\bibnamefont {Zhu}},\ and\ \bibinfo {author}
		{\bibfnamefont {Y.}~\bibnamefont {Ma}},\ }\href
	{https://doi.org/https://doi.org/10.1016/j.cpc.2012.05.008} {\bibfield
		{journal} {\bibinfo  {journal} {Comput. Phys. Commun.}\ }\textbf {\bibinfo
			{volume} {183}},\ \bibinfo {pages} {2063} (\bibinfo {year}
		{2012}{\natexlab{a}})}\BibitemShut {NoStop}%
	\bibitem [{\citenamefont {Wang}\ \emph
		{et~al.}(2012{\natexlab{b}})\citenamefont {Wang}, \citenamefont {Miao},
		\citenamefont {Lv}, \citenamefont {Zhu}, \citenamefont {Yin}, \citenamefont
		{Liu},\ and\ \citenamefont {Ma}}]{CALYPSO_2D}%
	\BibitemOpen
	\bibfield  {author} {\bibinfo {author} {\bibfnamefont {Y.}~\bibnamefont
			{Wang}}, \bibinfo {author} {\bibfnamefont {M.}~\bibnamefont {Miao}}, \bibinfo
		{author} {\bibfnamefont {J.}~\bibnamefont {Lv}}, \bibinfo {author}
		{\bibfnamefont {L.}~\bibnamefont {Zhu}}, \bibinfo {author} {\bibfnamefont
			{K.}~\bibnamefont {Yin}}, \bibinfo {author} {\bibfnamefont {H.}~\bibnamefont
			{Liu}},\ and\ \bibinfo {author} {\bibfnamefont {Y.}~\bibnamefont {Ma}},\
	}\href {https://doi.org/10.1063/1.4769731} {\bibfield  {journal} {\bibinfo
			{journal} {J. Chem. Phys.}\ }\textbf {\bibinfo {volume} {137}},\ \bibinfo
		{pages} {224108} (\bibinfo {year} {2012}{\natexlab{b}})}\BibitemShut
	{NoStop}%
	\bibitem [{\citenamefont {Hohenberg}\ and\ \citenamefont {Kohn}(1964)}]{DFT_1}%
	\BibitemOpen
	\bibfield  {author} {\bibinfo {author} {\bibfnamefont {P.}~\bibnamefont
			{Hohenberg}}\ and\ \bibinfo {author} {\bibfnamefont {W.}~\bibnamefont
			{Kohn}},\ }\href {https://doi.org/10.1103/PhysRev.136.B864} {\bibfield
		{journal} {\bibinfo  {journal} {Phys. Rev.}\ }\textbf {\bibinfo {volume}
			{136}},\ \bibinfo {pages} {B864} (\bibinfo {year} {1964})}\BibitemShut
	{NoStop}%
	\bibitem [{\citenamefont {Kohn}\ and\ \citenamefont {Sham}(1965)}]{DFT_2}%
	\BibitemOpen
	\bibfield  {author} {\bibinfo {author} {\bibfnamefont {W.}~\bibnamefont
			{Kohn}}\ and\ \bibinfo {author} {\bibfnamefont {L.~J.}\ \bibnamefont
			{Sham}},\ }\href {https://doi.org/10.1103/PhysRev.140.A1133} {\bibfield
		{journal} {\bibinfo  {journal} {Phys. Rev.}\ }\textbf {\bibinfo {volume}
			{140}},\ \bibinfo {pages} {A1133} (\bibinfo {year} {1965})}\BibitemShut
	{NoStop}%
	\bibitem [{\citenamefont {Xie}\ and\ \citenamefont
		{Grossman}(2018{\natexlab{a}})}]{CGCNN}%
	\BibitemOpen
	\bibfield  {author} {\bibinfo {author} {\bibfnamefont {T.}~\bibnamefont
			{Xie}}\ and\ \bibinfo {author} {\bibfnamefont {J.~C.}\ \bibnamefont
			{Grossman}},\ }\href {https://doi.org/10.1103/PhysRevLett.120.145301}
	{\bibfield  {journal} {\bibinfo  {journal} {Phys. Rev. Lett.}\ }\textbf
		{\bibinfo {volume} {120}},\ \bibinfo {pages} {145301} (\bibinfo {year}
		{2018}{\natexlab{a}})}\BibitemShut {NoStop}%
	\bibitem [{\citenamefont {Choudhary}\ and\ \citenamefont
		{DeCost}(2021)}]{ALIGNN}%
	\BibitemOpen
	\bibfield  {author} {\bibinfo {author} {\bibfnamefont {K.}~\bibnamefont
			{Choudhary}}\ and\ \bibinfo {author} {\bibfnamefont {B.}~\bibnamefont
			{DeCost}},\ }\href {https://doi.org/10.1038/s41524-021-00650-1} {\bibfield
		{journal} {\bibinfo  {journal} {npj Comput. Mater.}\ }\textbf {\bibinfo
			{volume} {7}},\ \bibinfo {pages} {185} (\bibinfo {year} {2021})}\BibitemShut
	{NoStop}%
	\bibitem [{\citenamefont {Chen}\ \emph {et~al.}(2019)\citenamefont {Chen},
		\citenamefont {Ye}, \citenamefont {Zuo}, \citenamefont {Zheng},\ and\
		\citenamefont {Ong}}]{MEGNet}%
	\BibitemOpen
	\bibfield  {author} {\bibinfo {author} {\bibfnamefont {C.}~\bibnamefont
			{Chen}}, \bibinfo {author} {\bibfnamefont {W.}~\bibnamefont {Ye}}, \bibinfo
		{author} {\bibfnamefont {Y.}~\bibnamefont {Zuo}}, \bibinfo {author}
		{\bibfnamefont {C.}~\bibnamefont {Zheng}},\ and\ \bibinfo {author}
		{\bibfnamefont {S.~P.}\ \bibnamefont {Ong}},\ }\href
	{https://doi.org/10.1021/acs.chemmater.9b01294} {\bibfield  {journal}
		{\bibinfo  {journal} {Chem. Mater.}\ }\textbf {\bibinfo {volume} {31}},\
		\bibinfo {pages} {3564} (\bibinfo {year} {2019})}\BibitemShut {NoStop}%
	\bibitem [{\citenamefont {Gilmer}\ \emph {et~al.}(2017)\citenamefont {Gilmer},
		\citenamefont {Schoenholz}, \citenamefont {Riley}, \citenamefont {Vinyals},\
		and\ \citenamefont {Dahl}}]{MPNN}%
	\BibitemOpen
	\bibfield  {author} {\bibinfo {author} {\bibfnamefont {J.}~\bibnamefont
			{Gilmer}}, \bibinfo {author} {\bibfnamefont {S.~S.}\ \bibnamefont
			{Schoenholz}}, \bibinfo {author} {\bibfnamefont {P.~F.}\ \bibnamefont
			{Riley}}, \bibinfo {author} {\bibfnamefont {O.}~\bibnamefont {Vinyals}},\
		and\ \bibinfo {author} {\bibfnamefont {G.~E.}\ \bibnamefont {Dahl}},\ }in\
	\href@noop {} {\emph {\bibinfo {booktitle} {International conference on
				machine learning}}}\ (\bibinfo {organization} {PMLR},\ \bibinfo {year}
	{2017})\ pp.\ \bibinfo {pages} {1263--1272}\BibitemShut {NoStop}%
	\bibitem [{\citenamefont {Xie}\ and\ \citenamefont
		{Grossman}(2018{\natexlab{b}})}]{CGCNN_XAI}%
	\BibitemOpen
	\bibfield  {author} {\bibinfo {author} {\bibfnamefont {T.}~\bibnamefont
			{Xie}}\ and\ \bibinfo {author} {\bibfnamefont {J.~C.}\ \bibnamefont
			{Grossman}},\ }\href {https://doi.org/10.1063/1.5047803} {\bibfield
		{journal} {\bibinfo  {journal} {J. Chem. Phys.}\ }\textbf {\bibinfo {volume}
			{149}},\ \bibinfo {pages} {174111} (\bibinfo {year}
		{2018}{\natexlab{b}})}\BibitemShut {NoStop}%
	\bibitem [{\citenamefont {Hsu}\ \emph {et~al.}(2022)\citenamefont {Hsu},
		\citenamefont {Pham}, \citenamefont {Keilbart} \emph {et~al.}}]{ALIGNN-d}%
	\BibitemOpen
	\bibfield  {author} {\bibinfo {author} {\bibfnamefont {T.}~\bibnamefont
			{Hsu}}, \bibinfo {author} {\bibfnamefont {T.~A.}\ \bibnamefont {Pham}},
		\bibinfo {author} {\bibfnamefont {N.}~\bibnamefont {Keilbart}}, \emph
		{et~al.},\ }\href {https://doi.org/10.1038/s41524-022-00841-4} {\bibfield
		{journal} {\bibinfo  {journal} {npj Comput. Mater.}\ }\textbf {\bibinfo
			{volume} {8}},\ \bibinfo {pages} {151} (\bibinfo {year} {2022})}\BibitemShut
	{NoStop}%
	\bibitem [{\citenamefont {Mardt}\ \emph {et~al.}(2018)\citenamefont {Mardt},
		\citenamefont {Pasquali}, \citenamefont {Wu},\ and\ \citenamefont
		{Noé}}]{VAMPnet}%
	\BibitemOpen
	\bibfield  {author} {\bibinfo {author} {\bibfnamefont {A.}~\bibnamefont
			{Mardt}}, \bibinfo {author} {\bibfnamefont {L.}~\bibnamefont {Pasquali}},
		\bibinfo {author} {\bibfnamefont {H.}~\bibnamefont {Wu}},\ and\ \bibinfo
		{author} {\bibfnamefont {F.}~\bibnamefont {Noé}},\ }\href@noop {} {\bibfield
		{journal} {\bibinfo  {journal} {Nat Commun}\ }\textbf {\bibinfo {volume}
			{9}},\ \bibinfo {pages} {5} (\bibinfo {year} {2018})}\BibitemShut {NoStop}%
	\bibitem [{\citenamefont {Xie}\ \emph {et~al.}(2019)\citenamefont {Xie},
		\citenamefont {France-Lanord}, \citenamefont {Wang}, \citenamefont
		{Shao-Horn},\ and\ \citenamefont {Grossman}}]{GDNet}%
	\BibitemOpen
	\bibfield  {author} {\bibinfo {author} {\bibfnamefont {T.}~\bibnamefont
			{Xie}}, \bibinfo {author} {\bibfnamefont {A.}~\bibnamefont {France-Lanord}},
		\bibinfo {author} {\bibfnamefont {Y.}~\bibnamefont {Wang}}, \bibinfo {author}
		{\bibfnamefont {Y.}~\bibnamefont {Shao-Horn}},\ and\ \bibinfo {author}
		{\bibfnamefont {J.~C.}\ \bibnamefont {Grossman}},\ }\href@noop {} {\bibfield
		{journal} {\bibinfo  {journal} {Nat Commun}\ }\textbf {\bibinfo {volume}
			{10}} (\bibinfo {year} {2019})}\BibitemShut {NoStop}%
	\bibitem [{\citenamefont {Fan}\ \emph {et~al.}(2021)\citenamefont {Fan},
		\citenamefont {Yan}, \citenamefont {Tripp} \emph {et~al.}}]{Biphenylene1}%
	\BibitemOpen
	\bibfield  {author} {\bibinfo {author} {\bibfnamefont {Q.}~\bibnamefont
			{Fan}}, \bibinfo {author} {\bibfnamefont {L.}~\bibnamefont {Yan}}, \bibinfo
		{author} {\bibfnamefont {M.~W.}\ \bibnamefont {Tripp}}, \emph {et~al.},\
	}\href {https://doi.org/10.1126/science.abg4509} {\bibfield  {journal}
		{\bibinfo  {journal} {Science}\ }\textbf {\bibinfo {volume} {372}},\ \bibinfo
		{pages} {852} (\bibinfo {year} {2021})}\BibitemShut {NoStop}%
	\bibitem [{\citenamefont {Sheng}\ \emph {et~al.}(2011)\citenamefont {Sheng},
		\citenamefont {Yan}, \citenamefont {Ye}, \citenamefont {Zheng},\ and\
		\citenamefont {Su}}]{2011-PRL-Tcarbon}%
	\BibitemOpen
	\bibfield  {author} {\bibinfo {author} {\bibfnamefont {X.-L.}\ \bibnamefont
			{Sheng}}, \bibinfo {author} {\bibfnamefont {Q.-B.}\ \bibnamefont {Yan}},
		\bibinfo {author} {\bibfnamefont {F.}~\bibnamefont {Ye}}, \bibinfo {author}
		{\bibfnamefont {Q.-R.}\ \bibnamefont {Zheng}},\ and\ \bibinfo {author}
		{\bibfnamefont {G.}~\bibnamefont {Su}},\ }\href
	{https://doi.org/10.1103/PhysRevLett.106.155703} {\bibfield  {journal}
		{\bibinfo  {journal} {Phys. Rev. Lett.}\ }\textbf {\bibinfo {volume} {106}},\
		\bibinfo {pages} {155703} (\bibinfo {year} {2011})}\BibitemShut {NoStop}%
	\bibitem [{\citenamefont {Zhang}\ \emph {et~al.}(2017)\citenamefont {Zhang},
		\citenamefont {Wang}, \citenamefont {Zhu} \emph
		{et~al.}}]{2017-NC-T-carbon-exp}%
	\BibitemOpen
	\bibfield  {author} {\bibinfo {author} {\bibfnamefont {J.}~\bibnamefont
			{Zhang}}, \bibinfo {author} {\bibfnamefont {R.}~\bibnamefont {Wang}},
		\bibinfo {author} {\bibfnamefont {X.}~\bibnamefont {Zhu}}, \emph {et~al.},\
	}\href {https://doi.org/10.1038/s41467-017-00817-9} {\bibfield  {journal}
		{\bibinfo  {journal} {Nat Commun}\ }\textbf {\bibinfo {volume} {8}},\
		\bibinfo {pages} {683} (\bibinfo {year} {2017})}\BibitemShut {NoStop}%
	\bibitem [{\citenamefont {Hudspeth}\ \emph {et~al.}(2010)\citenamefont
		{Hudspeth}, \citenamefont {Whitman}, \citenamefont {Barone},\ and\
		\citenamefont {Peralta}}]{BCN_1}%
	\BibitemOpen
	\bibfield  {author} {\bibinfo {author} {\bibfnamefont {M.~A.}\ \bibnamefont
			{Hudspeth}}, \bibinfo {author} {\bibfnamefont {B.~W.}\ \bibnamefont
			{Whitman}}, \bibinfo {author} {\bibfnamefont {V.}~\bibnamefont {Barone}},\
		and\ \bibinfo {author} {\bibfnamefont {J.~E.}\ \bibnamefont {Peralta}},\
	}\href {https://doi.org/doi: 10.1021/nn100758h} {\bibfield  {journal}
		{\bibinfo  {journal} {ACS Nano}\ }\textbf {\bibinfo {volume} {4}},\ \bibinfo
		{pages} {4565} (\bibinfo {year} {2010})}\BibitemShut {NoStop}%
	\bibitem [{\citenamefont {Demirci}\ \emph {et~al.}(2022)\citenamefont
		{Demirci}, \citenamefont {\ifmmode \mbox{\c{C}}\else \c{C}\fi{}all\ifmmode
			\imath \else \i \fi{}o\ifmmode~\breve{g}\else \u{g}\fi{}lu}, \citenamefont
		{G\"orkan}, \citenamefont {Akt\"urk},\ and\ \citenamefont
		{Ciraci}}]{Biphenylene2}%
	\BibitemOpen
	\bibfield  {author} {\bibinfo {author} {\bibfnamefont {S.}~\bibnamefont
			{Demirci}}, \bibinfo {author} {\bibfnamefont {i.~m.~c.}\ \bibnamefont
			{\ifmmode \mbox{\c{C}}\else \c{C}\fi{}all\ifmmode \imath \else \i
				\fi{}o\ifmmode~\breve{g}\else \u{g}\fi{}lu}}, \bibinfo {author}
		{\bibfnamefont {T.}~\bibnamefont {G\"orkan}}, \bibinfo {author}
		{\bibfnamefont {E.}~\bibnamefont {Akt\"urk}},\ and\ \bibinfo {author}
		{\bibfnamefont {S.}~\bibnamefont {Ciraci}},\ }\href
	{https://doi.org/10.1103/PhysRevB.105.035408} {\bibfield  {journal} {\bibinfo
			{journal} {Phys. Rev. B}\ }\textbf {\bibinfo {volume} {105}},\ \bibinfo
		{pages} {035408} (\bibinfo {year} {2022})}\BibitemShut {NoStop}%
	\bibitem [{\citenamefont {Liang}\ \emph {et~al.}(2021)\citenamefont {Liang},
		\citenamefont {Zhong}, \citenamefont {Huang},\ and\ \citenamefont
		{Duan}}]{Liang_1}%
	\BibitemOpen
	\bibfield  {author} {\bibinfo {author} {\bibfnamefont {H.}~\bibnamefont
			{Liang}}, \bibinfo {author} {\bibfnamefont {H.}~\bibnamefont {Zhong}},
		\bibinfo {author} {\bibfnamefont {S.}~\bibnamefont {Huang}},\ and\ \bibinfo
		{author} {\bibfnamefont {Y.}~\bibnamefont {Duan}},\ }\href
	{https://doi.org/10.1021/acs.jpclett.1c03248} {\bibfield  {journal} {\bibinfo
			{journal} {J. Phys. Chem. Lett.}\ }\textbf {\bibinfo {volume} {14}},\
		\bibinfo {pages} {10975} (\bibinfo {year} {2021})}\BibitemShut {NoStop}%
	\bibitem [{\citenamefont {Bafekry}\ \emph {et~al.}(2019)\citenamefont
		{Bafekry}, \citenamefont {Shayesteh},\ and\ \citenamefont
		{Peeters}}]{CN_Study}%
	\BibitemOpen
	\bibfield  {author} {\bibinfo {author} {\bibfnamefont {A.}~\bibnamefont
			{Bafekry}}, \bibinfo {author} {\bibfnamefont {S.~F.}\ \bibnamefont
			{Shayesteh}},\ and\ \bibinfo {author} {\bibfnamefont {F.~M.}\ \bibnamefont
			{Peeters}},\ }\href {https://doi.org/10.1063/1.5120525} {\bibfield  {journal}
		{\bibinfo  {journal} {J. Appl. Phys.}\ }\textbf {\bibinfo {volume} {126}},\
		\bibinfo {pages} {215104} (\bibinfo {year} {2019})}\BibitemShut {NoStop}%
	\bibitem [{\citenamefont {Luo}\ \emph {et~al.}(2011)\citenamefont {Luo},
		\citenamefont {Yang}, \citenamefont {Liu} \emph {et~al.}}]{BC_PSO_Search}%
	\BibitemOpen
	\bibfield  {author} {\bibinfo {author} {\bibfnamefont {X.}~\bibnamefont
			{Luo}}, \bibinfo {author} {\bibfnamefont {J.}~\bibnamefont {Yang}}, \bibinfo
		{author} {\bibfnamefont {H.}~\bibnamefont {Liu}}, \emph {et~al.},\ }\href
	{https://doi.org/10.1021/ja2072753} {\bibfield  {journal} {\bibinfo
			{journal} {J. Am. Chem. Soc.}\ }\textbf {\bibinfo {volume} {133}},\ \bibinfo
		{pages} {16285} (\bibinfo {year} {2011})}\BibitemShut {NoStop}%
	\bibitem [{\citenamefont {Zhou}\ \emph {et~al.}(2021)\citenamefont {Zhou},
		\citenamefont {Chen}, \citenamefont {Shu} \emph {et~al.}}]{BN_PSO_Search}%
	\BibitemOpen
	\bibfield  {author} {\bibinfo {author} {\bibfnamefont {X.}~\bibnamefont
			{Zhou}}, \bibinfo {author} {\bibfnamefont {X.}~\bibnamefont {Chen}}, \bibinfo
		{author} {\bibfnamefont {C.}~\bibnamefont {Shu}}, \emph {et~al.},\ }\href
	{https://doi.org/10.1021/acsami.1c08331} {\bibfield  {journal} {\bibinfo
			{journal} {ACS Appl. Mater. Interfaces}\ }\textbf {\bibinfo {volume} {13}},\
		\bibinfo {pages} {41169} (\bibinfo {year} {2021})}\BibitemShut {NoStop}%
	\bibitem [{\citenamefont {Hahn}\ \emph {et~al.}(1984)\citenamefont {Hahn},
		\citenamefont {Shmueli},\ and\ \citenamefont {Wilson}}]{Wyckoff}%
	\BibitemOpen
	\bibfield  {author} {\bibinfo {author} {\bibfnamefont {T.}~\bibnamefont
			{Hahn}}, \bibinfo {author} {\bibfnamefont {U.}~\bibnamefont {Shmueli}},\ and\
		\bibinfo {author} {\bibfnamefont {A.}~\bibnamefont {Wilson}},\ }\href
	{https://doi.org/10.1107/97809553602060000114} {\emph {\bibinfo {title}
			{International tables for crystallography}}}\ (\bibinfo  {publisher} {D.
		Reidel Pub. Co.; Sold and distributed in the USA and Canada by Kluwer
		Academic Publishers Group},\ \bibinfo {year} {1984})\BibitemShut {NoStop}%
	\bibitem [{\citenamefont {Oganov}\ and\ \citenamefont
		{Glass}(2006)}]{Gridding}%
	\BibitemOpen
	\bibfield  {author} {\bibinfo {author} {\bibfnamefont {A.~R.}\ \bibnamefont
			{Oganov}}\ and\ \bibinfo {author} {\bibfnamefont {C.~W.}\ \bibnamefont
			{Glass}},\ }\href {https://doi.org/10.1063/1.2210932} {\bibfield  {journal}
		{\bibinfo  {journal} {J. Chem. Phys.}\ }\textbf {\bibinfo {volume} {124}},\
		\bibinfo {pages} {244704} (\bibinfo {year} {2006})}\BibitemShut {NoStop}%
	\bibitem [{\citenamefont {Rasmussen}\ and\ \citenamefont
		{Williams}(2006)}]{Gaussian_Model}%
	\BibitemOpen
	\bibfield  {author} {\bibinfo {author} {\bibfnamefont {C.~E.}\ \bibnamefont
			{Rasmussen}}\ and\ \bibinfo {author} {\bibfnamefont {C.~K.~I.}\ \bibnamefont
			{Williams}},\ }\href {https://doi.org/10.7551/mitpress/3206.003.0010} {\emph
		{\bibinfo {title} {Gaussian Processes for Machine Learning}}}\ (\bibinfo
	{publisher} {MIT Press},\ \bibinfo {year} {2006})\BibitemShut {NoStop}%
	\bibitem [{\citenamefont {Shahriari}\ \emph {et~al.}(2016)\citenamefont
		{Shahriari}, \citenamefont {Swersky}, \citenamefont {Wang}, \citenamefont
		{Adams},\ and\ \citenamefont {de~Freitas}}]{Bayesian}%
	\BibitemOpen
	\bibfield  {author} {\bibinfo {author} {\bibfnamefont {B.}~\bibnamefont
			{Shahriari}}, \bibinfo {author} {\bibfnamefont {K.}~\bibnamefont {Swersky}},
		\bibinfo {author} {\bibfnamefont {Z.}~\bibnamefont {Wang}}, \bibinfo {author}
		{\bibfnamefont {R.~P.}\ \bibnamefont {Adams}},\ and\ \bibinfo {author}
		{\bibfnamefont {N.}~\bibnamefont {de~Freitas}},\ }\href@noop {} {\bibfield
		{journal} {\bibinfo  {journal} {Proceedings of the IEEE}\ }\textbf {\bibinfo
			{volume} {104}},\ \bibinfo {pages} {148} (\bibinfo {year}
		{2016})}\BibitemShut {NoStop}%
	\bibitem [{\citenamefont {Choudhary}\ \emph {et~al.}(2020)\citenamefont
		{Choudhary}, \citenamefont {Garrity}, \citenamefont {Reid} \emph
		{et~al.}}]{JARVIS}%
	\BibitemOpen
	\bibfield  {author} {\bibinfo {author} {\bibfnamefont {K.}~\bibnamefont
			{Choudhary}}, \bibinfo {author} {\bibfnamefont {K.~F.}\ \bibnamefont
			{Garrity}}, \bibinfo {author} {\bibfnamefont {A.~C.~E.}\ \bibnamefont
			{Reid}}, \emph {et~al.},\ }\href {https://doi.org/10.1038/s41524-020-00440-1}
	{\bibfield  {journal} {\bibinfo  {journal} {npj Comput. Mater.}\ }\textbf
		{\bibinfo {volume} {6}},\ \bibinfo {pages} {173} (\bibinfo {year}
		{2020})}\BibitemShut {NoStop}%
	\bibitem [{\citenamefont {Haastrup}\ \emph {et~al.}(2018)\citenamefont
		{Haastrup}, \citenamefont {Strange}, \citenamefont {Pandey} \emph
		{et~al.}}]{C2DB}%
	\BibitemOpen
	\bibfield  {author} {\bibinfo {author} {\bibfnamefont {S.}~\bibnamefont
			{Haastrup}}, \bibinfo {author} {\bibfnamefont {M.}~\bibnamefont {Strange}},
		\bibinfo {author} {\bibfnamefont {M.}~\bibnamefont {Pandey}}, \emph
		{et~al.},\ }\href {https://doi.org/10.1088/2053-1583/aacfc1} {\bibfield
		{journal} {\bibinfo  {journal} {2D Mater.}\ }\textbf {\bibinfo {volume}
			{5}},\ \bibinfo {pages} {042002} (\bibinfo {year} {2018})}\BibitemShut
	{NoStop}%
	\bibitem [{\citenamefont {Zhou}\ \emph {et~al.}(2019)\citenamefont {Zhou},
		\citenamefont {Shen}, \citenamefont {Costa} \emph {et~al.}}]{2DMatPedia}%
	\BibitemOpen
	\bibfield  {author} {\bibinfo {author} {\bibfnamefont {J.}~\bibnamefont
			{Zhou}}, \bibinfo {author} {\bibfnamefont {L.}~\bibnamefont {Shen}}, \bibinfo
		{author} {\bibfnamefont {M.~D.}\ \bibnamefont {Costa}}, \emph {et~al.},\
	}\href {https://doi.org/10.1038/s41597-019-0097-3} {\bibfield  {journal}
		{\bibinfo  {journal} {Sci. Data}\ }\textbf {\bibinfo {volume} {6}},\ \bibinfo
		{pages} {86} (\bibinfo {year} {2019})}\BibitemShut {NoStop}%
	\bibitem [{\citenamefont {Kingma}\ and\ \citenamefont {Ba}(2015)}]{Adam}%
	\BibitemOpen
	\bibfield  {author} {\bibinfo {author} {\bibfnamefont {D.~P.}\ \bibnamefont
			{Kingma}}\ and\ \bibinfo {author} {\bibfnamefont {J.}~\bibnamefont {Ba}},\
	}in\ \href@noop {} {\emph {\bibinfo {booktitle} {International Conference on
				Learning Representations}}}\ (\bibinfo {year} {2015})\BibitemShut {NoStop}%
	\bibitem [{\citenamefont {Weiss}\ \emph {et~al.}(2016)\citenamefont {Weiss},
		\citenamefont {Khoshgoftaar},\ and\ \citenamefont
		{Wang}}]{Transfer_Learning}%
	\BibitemOpen
	\bibfield  {author} {\bibinfo {author} {\bibfnamefont {K.}~\bibnamefont
			{Weiss}}, \bibinfo {author} {\bibfnamefont {T.~M.}\ \bibnamefont
			{Khoshgoftaar}},\ and\ \bibinfo {author} {\bibfnamefont {D.}~\bibnamefont
			{Wang}},\ }\href {https://doi.org/10.1186/s40537-016-0043-6} {\bibfield
		{journal} {\bibinfo  {journal} {J. Big Data}\ }\textbf {\bibinfo {volume}
			{3}},\ \bibinfo {pages} {9} (\bibinfo {year} {2016})}\BibitemShut {NoStop}%
	\bibitem [{\citenamefont {Laurens}\ and\ \citenamefont {Hinton}(2008)}]{TSNE}%
	\BibitemOpen
	\bibfield  {author} {\bibinfo {author} {\bibfnamefont {V.~D.~M.}\
			\bibnamefont {Laurens}}\ and\ \bibinfo {author} {\bibfnamefont
			{G.}~\bibnamefont {Hinton}},\ }\href@noop {} {\bibfield  {journal} {\bibinfo
			{journal} {J. Mach. Learn. Res.}\ }\textbf {\bibinfo {volume} {9}},\ \bibinfo
		{pages} {2579} (\bibinfo {year} {2008})}\BibitemShut {NoStop}%
	\bibitem [{\citenamefont {Bahmani}\ \emph {et~al.}(2012)\citenamefont
		{Bahmani}, \citenamefont {Moseley}, \citenamefont {Vattani}, \citenamefont
		{Kumar},\ and\ \citenamefont {Vassilvitskii}}]{Kmeans}%
	\BibitemOpen
	\bibfield  {author} {\bibinfo {author} {\bibfnamefont {B.}~\bibnamefont
			{Bahmani}}, \bibinfo {author} {\bibfnamefont {B.}~\bibnamefont {Moseley}},
		\bibinfo {author} {\bibfnamefont {A.}~\bibnamefont {Vattani}}, \bibinfo
		{author} {\bibfnamefont {R.}~\bibnamefont {Kumar}},\ and\ \bibinfo {author}
		{\bibfnamefont {S.}~\bibnamefont {Vassilvitskii}},\ }\href@noop {} {\bibfield
		{journal} {\bibinfo  {journal} {Proceedings of the VLDB Endowment}\ }\textbf
		{\bibinfo {volume} {5}} (\bibinfo {year} {2012})}\BibitemShut {NoStop}%
	\bibitem [{\citenamefont {Jiménez-Luna}\ \emph {et~al.}(2020)\citenamefont
		{Jiménez-Luna}, \citenamefont {Grisoni},\ and\ \citenamefont
		{Schneider}}]{XAI_Add_Model_Review}%
	\BibitemOpen
	\bibfield  {author} {\bibinfo {author} {\bibfnamefont {J.}~\bibnamefont
			{Jiménez-Luna}}, \bibinfo {author} {\bibfnamefont {F.}~\bibnamefont
			{Grisoni}},\ and\ \bibinfo {author} {\bibfnamefont {G.}~\bibnamefont
			{Schneider}},\ }\href {https://doi.org/10.1038/s42256-020-00236-4} {\bibfield
		{journal} {\bibinfo  {journal} {Nat. Mach. Intell.}\ }\textbf {\bibinfo
			{volume} {2}},\ \bibinfo {pages} {573} (\bibinfo {year} {2020})}\BibitemShut
	{NoStop}%
	\bibitem [{\citenamefont {Kresse}\ and\ \citenamefont {Hafner}(1993)}]{VASP_1}%
	\BibitemOpen
	\bibfield  {author} {\bibinfo {author} {\bibfnamefont {G.}~\bibnamefont
			{Kresse}}\ and\ \bibinfo {author} {\bibfnamefont {J.}~\bibnamefont
			{Hafner}},\ }\href {https://doi.org/10.1103/PhysRevB.47.558} {\bibfield
		{journal} {\bibinfo  {journal} {Phys. Rev. B}\ }\textbf {\bibinfo {volume}
			{47}},\ \bibinfo {pages} {558} (\bibinfo {year} {1993})}\BibitemShut
	{NoStop}%
	\bibitem [{\citenamefont {Kresse}\ and\ \citenamefont {Hafner}(1994)}]{VASP_2}%
	\BibitemOpen
	\bibfield  {author} {\bibinfo {author} {\bibfnamefont {G.}~\bibnamefont
			{Kresse}}\ and\ \bibinfo {author} {\bibfnamefont {J.}~\bibnamefont
			{Hafner}},\ }\href {https://doi.org/10.1103/PhysRevB.49.14251} {\bibfield
		{journal} {\bibinfo  {journal} {Phys. Rev. B}\ }\textbf {\bibinfo {volume}
			{49}},\ \bibinfo {pages} {14251} (\bibinfo {year} {1994})}\BibitemShut
	{NoStop}%
	\bibitem [{\citenamefont {Kresse}\ and\ \citenamefont
		{Furthm\"uller}(1996)}]{VASP_3}%
	\BibitemOpen
	\bibfield  {author} {\bibinfo {author} {\bibfnamefont {G.}~\bibnamefont
			{Kresse}}\ and\ \bibinfo {author} {\bibfnamefont {J.}~\bibnamefont
			{Furthm\"uller}},\ }\href {https://doi.org/10.1103/PhysRevB.54.11169}
	{\bibfield  {journal} {\bibinfo  {journal} {Phys. Rev. B}\ }\textbf {\bibinfo
			{volume} {54}},\ \bibinfo {pages} {11169} (\bibinfo {year}
		{1996})}\BibitemShut {NoStop}%
	\bibitem [{\citenamefont {Perdew}\ \emph {et~al.}(1996)\citenamefont {Perdew},
		\citenamefont {Burke},\ and\ \citenamefont {Ernzerhof}}]{PBE}%
	\BibitemOpen
	\bibfield  {author} {\bibinfo {author} {\bibfnamefont {J.~P.}\ \bibnamefont
			{Perdew}}, \bibinfo {author} {\bibfnamefont {K.}~\bibnamefont {Burke}},\ and\
		\bibinfo {author} {\bibfnamefont {M.}~\bibnamefont {Ernzerhof}},\ }\href
	{https://doi.org/10.1103/PhysRevLett.77.3865} {\bibfield  {journal} {\bibinfo
			{journal} {Phys. Rev. Lett.}\ }\textbf {\bibinfo {volume} {77}},\ \bibinfo
		{pages} {3865} (\bibinfo {year} {1996})}\BibitemShut {NoStop}%
	\bibitem [{\citenamefont {Bl\"ochl}(1994)}]{PAW_1}%
	\BibitemOpen
	\bibfield  {author} {\bibinfo {author} {\bibfnamefont {P.~E.}\ \bibnamefont
			{Bl\"ochl}},\ }\href {https://doi.org/10.1103/PhysRevB.50.17953} {\bibfield
		{journal} {\bibinfo  {journal} {Phys. Rev. B}\ }\textbf {\bibinfo {volume}
			{50}},\ \bibinfo {pages} {17953} (\bibinfo {year} {1994})}\BibitemShut
	{NoStop}%
	\bibitem [{\citenamefont {Kresse}\ and\ \citenamefont {Joubert}(1999)}]{PAW_2}%
	\BibitemOpen
	\bibfield  {author} {\bibinfo {author} {\bibfnamefont {G.}~\bibnamefont
			{Kresse}}\ and\ \bibinfo {author} {\bibfnamefont {D.}~\bibnamefont
			{Joubert}},\ }\href {https://doi.org/10.1103/PhysRevB.59.1758} {\bibfield
		{journal} {\bibinfo  {journal} {Phys. Rev. B}\ }\textbf {\bibinfo {volume}
			{59}},\ \bibinfo {pages} {1758} (\bibinfo {year} {1999})}\BibitemShut
	{NoStop}%
	\bibitem [{\citenamefont {Heyd}\ \emph {et~al.}(2003)\citenamefont {Heyd},
		\citenamefont {Scuseria},\ and\ \citenamefont {Ernzerhof}}]{HSE06}%
	\BibitemOpen
	\bibfield  {author} {\bibinfo {author} {\bibfnamefont {J.}~\bibnamefont
			{Heyd}}, \bibinfo {author} {\bibfnamefont {G.~E.}\ \bibnamefont {Scuseria}},\
		and\ \bibinfo {author} {\bibfnamefont {M.}~\bibnamefont {Ernzerhof}},\ }\href
	{https://doi.org/10.1063/1.1564060} {\bibfield  {journal} {\bibinfo
			{journal} {J. Chem. Phys.}\ }\textbf {\bibinfo {volume} {118}},\ \bibinfo
		{pages} {8207} (\bibinfo {year} {2003})}\BibitemShut {NoStop}%
	\bibitem [{\citenamefont {Li}\ \emph {et~al.}(2014)\citenamefont {Li},
		\citenamefont {Carrete}, \citenamefont {{A. Katcho}},\ and\ \citenamefont
		{Mingo}}]{ShengBTE}%
	\BibitemOpen
	\bibfield  {author} {\bibinfo {author} {\bibfnamefont {W.}~\bibnamefont
			{Li}}, \bibinfo {author} {\bibfnamefont {J.}~\bibnamefont {Carrete}},
		\bibinfo {author} {\bibfnamefont {N.}~\bibnamefont {{A. Katcho}}},\ and\
		\bibinfo {author} {\bibfnamefont {N.}~\bibnamefont {Mingo}},\ }\href
	{https://doi.org/https://doi.org/10.1016/j.cpc.2014.02.015} {\bibfield
		{journal} {\bibinfo  {journal} {Comput. Phys. Commun.}\ }\textbf {\bibinfo
			{volume} {185}},\ \bibinfo {pages} {1747} (\bibinfo {year}
		{2014})}\BibitemShut {NoStop}%
	\bibitem [{\citenamefont {Adekoya}\ \emph {et~al.}(2020)\citenamefont
		{Adekoya}, \citenamefont {Qian}, \citenamefont {Gu} \emph
		{et~al.}}]{CN_Review}%
	\BibitemOpen
	\bibfield  {author} {\bibinfo {author} {\bibfnamefont {D.}~\bibnamefont
			{Adekoya}}, \bibinfo {author} {\bibfnamefont {S.}~\bibnamefont {Qian}},
		\bibinfo {author} {\bibfnamefont {X.}~\bibnamefont {Gu}}, \emph {et~al.},\
	}\href {https://doi.org/10.1007/s40820-020-00522-1} {\bibfield  {journal}
		{\bibinfo  {journal} {Nano-Micro Lett.}\ }\textbf {\bibinfo {volume} {13}},\
		\bibinfo {pages} {13} (\bibinfo {year} {2020})}\BibitemShut {NoStop}%
	\bibitem [{\citenamefont {Song}\ \emph {et~al.}(2012)\citenamefont {Song},
		\citenamefont {Liu}, \citenamefont {Reddy} \emph {et~al.}}]{BCN_Review_1}%
	\BibitemOpen
	\bibfield  {author} {\bibinfo {author} {\bibfnamefont {L.}~\bibnamefont
			{Song}}, \bibinfo {author} {\bibfnamefont {Z.}~\bibnamefont {Liu}}, \bibinfo
		{author} {\bibfnamefont {A.~L.~M.}\ \bibnamefont {Reddy}}, \emph {et~al.},\
	}\href {https://doi.org/https://doi.org/10.1002/adma.201201792} {\bibfield
		{journal} {\bibinfo  {journal} {Adv. Mater.}\ }\textbf {\bibinfo {volume}
			{24}},\ \bibinfo {pages} {4878} (\bibinfo {year} {2012})}\BibitemShut
	{NoStop}%
	\bibitem [{\citenamefont {Angizi}\ \emph {et~al.}(2020)\citenamefont {Angizi},
		\citenamefont {Akbar}, \citenamefont {Darestani-Farahani},\ and\
		\citenamefont {Kruse}}]{BCN_Review_2}%
	\BibitemOpen
	\bibfield  {author} {\bibinfo {author} {\bibfnamefont {S.}~\bibnamefont
			{Angizi}}, \bibinfo {author} {\bibfnamefont {M.~A.}\ \bibnamefont {Akbar}},
		\bibinfo {author} {\bibfnamefont {M.}~\bibnamefont {Darestani-Farahani}},\
		and\ \bibinfo {author} {\bibfnamefont {P.}~\bibnamefont {Kruse}},\ }\href
	{https://doi.org/10.1149/2162-8777/abb8ef} {\bibfield  {journal} {\bibinfo
			{journal} {{ECS} J. Solid State Sci. Technol.}\ }\textbf {\bibinfo {volume}
			{9}},\ \bibinfo {pages} {083004} (\bibinfo {year} {2020})}\BibitemShut
	{NoStop}%
	\bibitem [{\citenamefont {Ogitsu}\ \emph {et~al.}(2013)\citenamefont {Ogitsu},
		\citenamefont {Schwegler},\ and\ \citenamefont {Galli}}]{Boron_Deficient}%
	\BibitemOpen
	\bibfield  {author} {\bibinfo {author} {\bibfnamefont {T.}~\bibnamefont
			{Ogitsu}}, \bibinfo {author} {\bibfnamefont {E.}~\bibnamefont {Schwegler}},\
		and\ \bibinfo {author} {\bibfnamefont {G.}~\bibnamefont {Galli}},\ }\href
	{https://doi.org/10.1021/cr300356t} {\bibfield  {journal} {\bibinfo
			{journal} {Chem. Rev.}\ }\textbf {\bibinfo {volume} {113}},\ \bibinfo {pages}
		{3425} (\bibinfo {year} {2013})}\BibitemShut {NoStop}%
	\bibitem [{\citenamefont {Long}\ \emph {et~al.}(2019)\citenamefont {Long},
		\citenamefont {Wang}, \citenamefont {Fang},\ and\ \citenamefont
		{Hu}}]{Photodetectors1}%
	\BibitemOpen
	\bibfield  {author} {\bibinfo {author} {\bibfnamefont {M.}~\bibnamefont
			{Long}}, \bibinfo {author} {\bibfnamefont {P.}~\bibnamefont {Wang}}, \bibinfo
		{author} {\bibfnamefont {H.}~\bibnamefont {Fang}},\ and\ \bibinfo {author}
		{\bibfnamefont {W.}~\bibnamefont {Hu}},\ }\href
	{https://doi.org/https://doi.org/10.1002/adfm.201803807} {\bibfield
		{journal} {\bibinfo  {journal} {Adv. Funct. Mater.}\ }\textbf {\bibinfo
			{volume} {29}},\ \bibinfo {pages} {1803807} (\bibinfo {year}
		{2019})}\BibitemShut {NoStop}%
	\bibitem [{\citenamefont {Qiu}\ and\ \citenamefont
		{Huang}(2021)}]{Photodetectors2}%
	\BibitemOpen
	\bibfield  {author} {\bibinfo {author} {\bibfnamefont {Q.}~\bibnamefont
			{Qiu}}\ and\ \bibinfo {author} {\bibfnamefont {Z.}~\bibnamefont {Huang}},\
	}\href {https://doi.org/https://doi.org/10.1002/adma.202008126} {\bibfield
		{journal} {\bibinfo  {journal} {Adv. Mater.}\ }\textbf {\bibinfo {volume}
			{33}},\ \bibinfo {pages} {2008126} (\bibinfo {year} {2021})}\BibitemShut
	{NoStop}%
	\bibitem [{\citenamefont {Wicklein}\ \emph {et~al.}(2015)\citenamefont
		{Wicklein}, \citenamefont {Kocjan}, \citenamefont {Salazar-Alvarez} \emph
		{et~al.}}]{ThermalInsulator1}%
	\BibitemOpen
	\bibfield  {author} {\bibinfo {author} {\bibfnamefont {B.}~\bibnamefont
			{Wicklein}}, \bibinfo {author} {\bibfnamefont {A.}~\bibnamefont {Kocjan}},
		\bibinfo {author} {\bibfnamefont {G.}~\bibnamefont {Salazar-Alvarez}}, \emph
		{et~al.},\ }\href {https://doi.org/10.1038/nnano.2014.248} {\bibfield
		{journal} {\bibinfo  {journal} {Nature Nanotech}\ }\textbf {\bibinfo {volume}
			{10}},\ \bibinfo {pages} {277} (\bibinfo {year} {2015})}\BibitemShut
	{NoStop}%
	\bibitem [{\citenamefont {Si}\ \emph {et~al.}(2014)\citenamefont {Si},
		\citenamefont {Yu}, \citenamefont {Tang} \emph {et~al.}}]{ThermalInsulator2}%
	\BibitemOpen
	\bibfield  {author} {\bibinfo {author} {\bibfnamefont {Y.}~\bibnamefont
			{Si}}, \bibinfo {author} {\bibfnamefont {J.}~\bibnamefont {Yu}}, \bibinfo
		{author} {\bibfnamefont {X.}~\bibnamefont {Tang}}, \emph {et~al.},\ }\href
	{https://doi.org/10.1038/ncomms6802} {\bibfield  {journal} {\bibinfo
			{journal} {Nat Commun}\ }\textbf {\bibinfo {volume} {5}},\ \bibinfo {pages}
		{5802} (\bibinfo {year} {2014})}\BibitemShut {NoStop}%
	\bibitem [{\citenamefont {Biener}\ \emph {et~al.}(2011)\citenamefont {Biener},
		\citenamefont {Stadermann}, \citenamefont {Suss} \emph {et~al.}}]{Storage1}%
	\BibitemOpen
	\bibfield  {author} {\bibinfo {author} {\bibfnamefont {J.}~\bibnamefont
			{Biener}}, \bibinfo {author} {\bibfnamefont {M.}~\bibnamefont {Stadermann}},
		\bibinfo {author} {\bibfnamefont {M.}~\bibnamefont {Suss}}, \emph {et~al.},\
	}\href {https://doi.org/10.1039/C0EE00627K} {\bibfield  {journal} {\bibinfo
			{journal} {Energy Environ. Sci.}\ }\textbf {\bibinfo {volume} {4}},\ \bibinfo
		{pages} {656} (\bibinfo {year} {2011})}\BibitemShut {NoStop}%
	\bibitem [{\citenamefont {Hamedi}\ \emph {et~al.}(2013)\citenamefont {Hamedi},
		\citenamefont {Karabulut}, \citenamefont {Marais} \emph {et~al.}}]{Storage2}%
	\BibitemOpen
	\bibfield  {author} {\bibinfo {author} {\bibfnamefont {M.}~\bibnamefont
			{Hamedi}}, \bibinfo {author} {\bibfnamefont {E.}~\bibnamefont {Karabulut}},
		\bibinfo {author} {\bibfnamefont {A.}~\bibnamefont {Marais}}, \emph
		{et~al.},\ }\href {https://doi.org/https://doi.org/10.1002/anie.201305137}
	{\bibfield  {journal} {\bibinfo  {journal} {Angew. Chem. Int. Ed.}\ }\textbf
		{\bibinfo {volume} {52}},\ \bibinfo {pages} {12038} (\bibinfo {year}
		{2013})}\BibitemShut {NoStop}%
	\bibitem [{\citenamefont {Cheng}\ \emph {et~al.}(2022)\citenamefont {Cheng},
		\citenamefont {Gong},\ and\ \citenamefont {Yin}}]{GN_OA}%
	\BibitemOpen
	\bibfield  {author} {\bibinfo {author} {\bibfnamefont {G.}~\bibnamefont
			{Cheng}}, \bibinfo {author} {\bibfnamefont {X.-G.}\ \bibnamefont {Gong}},\
		and\ \bibinfo {author} {\bibfnamefont {W.-J.}\ \bibnamefont {Yin}},\ }\href
	{https://doi.org/10.1038/s41467-022-29241-4} {\bibfield  {journal} {\bibinfo
			{journal} {Nat Commun}\ }\textbf {\bibinfo {volume} {13}},\ \bibinfo {pages}
		{1492} (\bibinfo {year} {2022})}\BibitemShut {NoStop}%
	\bibitem [{\citenamefont {Zuo}\ \emph {et~al.}(2021)\citenamefont {Zuo},
		\citenamefont {Qin}, \citenamefont {Chen}, \citenamefont {Ye}, \citenamefont
		{Li}, \citenamefont {Luo},\ and\ \citenamefont {Ong}}]{BOWSR}%
	\BibitemOpen
	\bibfield  {author} {\bibinfo {author} {\bibfnamefont {Y.}~\bibnamefont
			{Zuo}}, \bibinfo {author} {\bibfnamefont {M.}~\bibnamefont {Qin}}, \bibinfo
		{author} {\bibfnamefont {C.}~\bibnamefont {Chen}}, \bibinfo {author}
		{\bibfnamefont {W.}~\bibnamefont {Ye}}, \bibinfo {author} {\bibfnamefont
			{X.}~\bibnamefont {Li}}, \bibinfo {author} {\bibfnamefont {J.}~\bibnamefont
			{Luo}},\ and\ \bibinfo {author} {\bibfnamefont {S.~P.}\ \bibnamefont {Ong}},\
	}\href {https://doi.org/https://doi.org/10.1016/j.mattod.2021.08.012}
	{\bibfield  {journal} {\bibinfo  {journal} {Mater. Today}\ }\textbf {\bibinfo
			{volume} {51}},\ \bibinfo {pages} {126} (\bibinfo {year} {2021})}\BibitemShut
	{NoStop}%
\end{thebibliography}

\end{document}